\documentclass[fleqn,usenatbib]{mnras}

\usepackage{newtxtext,newtxmath}

\usepackage[T1]{fontenc}

\DeclareRobustCommand{\VAN}[3]{#2}
\let\VANthebibliography\thebibliography
\def\thebibliography{\DeclareRobustCommand{\VAN}[3]{##3}\VANthebibliography}

\usepackage{graphicx}	
\usepackage{amsmath}	
\usepackage{booktabs}

\usepackage{enumitem} 

\newcommand{\MAGI}{{\tt MAGI}\,}
\newcommand{\GIZMO}{{\tt GIZMO}\,}
\newcommand{\FIRE}{{\tt FIRE-3}\,}

\newcommand{\eq}[1]{eq.~(\ref{eq:#1})}

\newcommand{\se}[1]{Section \ref{sec:#1}}

\newcommand{\Fig}[1]{Fig.~\ref{fig:#1}}
\newcommand{\Figs}[1]{Figs.~\ref{fig:#1}}
\newcommand{\Tab}[1]{Table~\ref{tab:#1}}
\newcommand{\be}{\begin{equation}}
\newcommand{\ee}{\end{equation}}
\newcommand{\bad}{\begin{equation} \begin{aligned}}
\newcommand{\ead}{\end{aligned} \end{equation}}

\newcommand{\Msun}{M_\odot}

\newcommand{\kpc}{\,{\rm kpc}}

\newcommand{\rhoc}{\rho_{\rm crit}}

\newcommand{\Mv}{M_{\rm vir}}
\newcommand{\Ms}{M_{\star}}

\newcommand{\fb}{f_{\rm b}}

\newcommand{\Rv}{R_{\rm vir}}

\newcommand{\rc}{r_{\rm c}}

\newcommand{\Vv}{V_{\rm vir}}

\newcommand{\jv}{j_{\rm vir}}
\newcommand{\Jv}{J_{\rm vir}}

\newcommand{\rmd}{{\rm d}}

\title[galaxy size regulators]{Controlled Experiments on Dark-Matter Halo Structure and Galaxy Morphology I: What Sets Galaxy Sizes?}

\author[G. Sun et al.]{
Guangze Sun,$^{1,2}$
Fangzhou Jiang,$^{2}$\thanks{Corresponding author: fangzhou.jiang@pku.edu.cn}
Jing Wang$^{2}$
\\
$^{1}$Department of Astronomy, School of Physics, Peking University, Beijing 100871, People’s Republic of China\\
$^{2}$Kavli Institute for Astronomy and Astrophysics, Peking University, Beijing 100871, People’s Republic of China\\
}

\date{Accepted 2026 May 25. Received 2026 May 06; in original form 2026 March 06}

\pubyear{\the\year{}}

\begin{document}
\label{firstpage}
\pagerange{\pageref{firstpage}--\pageref{lastpage}}
\maketitle

\begin{abstract}
The properties of galaxies are intricately linked to the characteristics of their host dark-matter haloes. 
We use a suite of controlled simulations of isolated galaxies to quantify how halo spin, concentration, inner density profile, and baryon fraction regulate galaxy sizes, at fixed halo mass of $\Mv=10^{11}\Msun$. 
We generate initial conditions of haloes and inhabitant spherical gas distributions in equilibrium, on a parameter grid spanned by these four halo parameters, and evolve the systems with the \GIZMO code and the \FIRE physics. 
The resulting half-mass radii of stars and cold baryons depend systematically on halo structure and baryon content: galaxy size increases with halo spin, decreases with halo concentration, is weakly sensitive to the inner density slope except in highly cuspy haloes, and is strongly suppressed at high baryon fractions.
We evaluate the relative importance of the halo parameters on galaxy size using different metrics including the quadratic response-surface method and random-forest regression, and consistently find halo concentration to be the most informative predictor of size. 
The baryon fraction shows a subtle, non-monotonic impact on size, by modulating how galaxy size depends on halo spin. 
Our results clarify which secondary parameters of host dark-matter haloes dominate the scatter in galaxy sizes at the massive-dwarf mass scale.
\end{abstract}




\section{Introduction}
In the standard $\Lambda$CDM paradigm of structure formation, dark matter (DM) dominates the mass budget and forms gravitationally bound structures, haloes, that provide gravitational potential wells in which galaxies form and evolve \citep[e.g.,][]{White78,Blumenthal84}. 
Theoretical models and numerical simulations based on this paradigm predict key statistical properties of the halo population, such as their mass function, spatial clustering, and internal structure \citep[e.g.,][]{Press74,Navarro96,Navarro97,Springel05b}.
Given these mature theories of DM haloes, a crucial step towards a theoretical framework of galaxy formation lies in establishing the galaxy-halo connection — the set of rules that govern how galaxies populate and evolve within these haloes \citep[see e.g.,][for a review]{Wechsler18}. 

A central question in the galaxy–halo connection is how the properties of dark-matter haloes regulate the morphology of their resident galaxies, particularly their characteristic sizes. 
Halo mass is widely recognized as the primary driver of galaxy morphology, a conclusion supported by several complementary lines of evidence. First, on average, galaxy size exhibits an approximately linear scaling with the virial radius of the host halo \citep[e.g.][]{Kravtsov13, Somerville18}, $R_{\rm gal} \simeq 0.02\,\Rv$. At a given epoch, the virial radius $\Rv$ is uniquely determined by halo mass and is commonly inferred from stellar masses using stellar mass–halo mass relations derived from abundance matching \citep[e.g.][]{Moster13, Behroozi13}.
Second, cosmological hydrodynamical simulations show that rotationally supported discs are most prominent over an intermediate mass range: lower-mass systems are more often kinematically hot or irregular, while the most massive galaxies tend to be spheroid dominated \citep[e.g.][]{Clauwens18,Tacchella19,Dekel20,RodriguezGomez22,Liang25a}. 
Observationally, at the slightly lower bright-dwarf scale, $\Mv \sim 10^{11}\Msun$, galaxies display striking morphological diversity, with stellar surface densities spanning more than two orders of magnitude and ultra-diffuse galaxies coexisting alongside compact dwarfs \citep[see e.g.][]{Sales22}. 
Thus, while halo mass sets the overall scale of galaxy size and morphology, the substantial scatter around these relations indicates that additional halo properties beyond mass must play an important role in regulating the final extent of galaxies.

The origin of this scatter remains debated. 
The angular momentum of the dark matter halo, often parameterized by the dimensionless spin parameter $\lambda$ \citep{Peebles69, Bullock01b}, has been considered a primary determinant of galaxy size. 
Classical models of galaxy formation, assuming conservation of specific angular momentum during baryonic collapse, predict that galaxy size scales with halo spin and virial radius, $R_{\rm{gal}}\propto\lambda R_{\rm{vir}}$ \citep[e.g.,][]{Fall80,Mo98}. 
However, hydrodynamical cosmological simulations have produced a mixed picture of how directly halo spin regulates galaxy size.

On the one hand, several studies show results in qualitative agreement with the angular-momentum-based models.
For example, in the Auriga zoom-in simulations of Milky Way-mass haloes, \citet{Grand17} found that disc scale length correlates with halo spin, with extended discs forming in haloes of high specific angular momentum.
Using the large-box TNG100 simulation, \citet{RodriguezGomez22} likewise showed that larger galaxies tend to reside in faster-spinning haloes (although the relation weakens toward the massive end and breaks down for $\Ms \gtrsim 10^{11}\Msun$, where accreted stars and spheroidal morphologies become important).
The strength of the size-spin correlation, however, varies across simulation suites.
For example, \citet{Yang23} compared four cosmological hydrodynamical simulation suites and showed that the size-spin relation is not universal: it is stronger in the IllustrisTNG/Auriga family, weaker in the EAGLE/APOSTLE family.
\citet{Liang25b,Liang25a} found in TNG50 that, while disc size at fixed halo mass is positively correlated with halo spin, it also exhibits significant anti-correlation with halo concentration and is affected by several other halo structural and environmental parameters.

On the other hand, some simulations exhibit weak correlations between halo spin and galaxy structure.
Focusing directly on galaxy spin and size, \citet{Jiang19} found from two suites of zoom-in simulations that galaxy spin is only weakly correlated with halo spin, and that the galaxy half-stellar-mass radius is better described by a concentration-corrected halo radius, $r_{1/2,\star} \propto c^{-0.7} \Rv$.
In the low-mass regime, central dwarf galaxies with stellar masses $\Ms \sim 10^{7\text{--}9}\Msun$ in the FIREbox simulation show little correlation between the scatter in galaxy half-stellar-mass radius at fixed halo virial radius and either halo spin or concentration \citep{Rohr22}, suggesting that baryonic processes dominate the size scatter in these systems.
\citet{Sales12} found that the rotational morphology of Milky Way-mass galaxies is only weakly related to halo spin or recent assembly history, and instead depends on the coherent alignment of baryonic angular momentum accreted over time.
Using the Illustris simulation, \citet{RodriguezGomez17} showed that halo spin correlates with rotational morphology mainly for dwarf galaxies and only weakly for galaxies above $\Ms \sim 10^{10}\Msun$.

Taken together, these studies indicate that the relevant size predictors depend on galaxy mass, morphology, size definition, and baryonic physics, leaving the primary drivers of galaxy size at fixed halo mass unsettled.

In addition to the mixed results from cosmological simulations, many of these studies share a conceptual limitation: the relations between galaxy size and host-halo structure are typically established using instantaneous halo properties, whereas galaxy sizes should, in principle, be more strongly influenced by the properties of their progenitor haloes at the time when the galaxy formed. 
Using cosmological simulations to connect galaxy properties to progenitor halo structure, however, presents two major challenges.
First, there is no consensus on how to define the halo “formation” time. 
Common definitions include the epoch when a halo first assembles half of its final mass \citep[e.g.][]{LaceyCole93, ShethTormen04}, the end of the fast accretion phase \citep[e.g.][]{Zhao03}, or the time of the last major merger \citep[e.g.][]{Li08}. 
It remains unclear which of these definitions most faithfully captures the physically relevant formation epoch for establishing galaxy–halo connections. 
Second, even if a specific definition is adopted, reconstructing the time evolution of halo structural parameters requires reprocessing particle-level data across all simulation snapshots, which is computationally and storage intensive.

Another limitation of analyses based on cosmological simulations is the restricted range of halo properties that they typically probe. For example, most simulations designed to study dwarf-galaxy structure employ zoom-in techniques or small simulation volumes \citep[e.g.][]{Wang15NIHAO, Hopkins18}. 
While these approaches achieve the numerical resolution required to resolve dwarf morphologies, they inevitably sacrifice statistical power and limit the diversity of halo structural properties.
This restriction poses a significant challenge for addressing questions such as how the inner density profile of a host dark-matter halo influences galaxy morphology. In the NIHAO and FIRE-2 simulations, haloes with masses around $10^{11}\Msun$ almost universally develop flat inner density cores \citep{Freundlich20, Lazar20}, whereas haloes in the IllustrisTNG simulations remain predominantly cuspy across a wide range of halo masses \citep{Bose19}. 

These considerations motivate complementing cosmological analyses with controlled numerical experiments designed to isolate specific physical dependencies. In controlled simulations, there is no ambiguity in the definition of formation time, as the properties of the system at formation are fully specified by the initial conditions. Moreover, halo parameters can be varied systematically, enabling a wide  baseline of halo structural properties by construction.

In this series of studies, we adopt idealized (isolated) galaxy simulations to identify which secondary dark-matter halo properties, beyond halo mass, most strongly regulate galaxy morphology. 
In this first paper of the series, we focus on a single halo mass, $\Mv = 10^{11}\Msun$, which corresponds to the characteristic halo mass of bright dwarf galaxies and the mass scale at which structural diversity is maximized \citep[e.g.][]{Cintio14, Oman15, Lazar24}. 
We systematically explore how variations in halo spin, concentration, inner density slope, and baryon fraction modulate galaxy size at fixed halo mass. 
In subsequent papers in this series, we plan to extend the analysis to a broader range of halo masses and examine the halo conditions governing additional aspects of galaxy morphology beyond size, such as clumpiness.

This paper is organized as follows. 
\se{method} describes the design of our parameter grid, the numerical implementation of the hydrodynamical simulations, and the analysis methods. 
\se{results} presents the correlations between galaxy size and initial halo properties and establishes a rank ordering of the relative importance of these halo parameters in regulating galaxy size.
In \se{discussion}, we discuss the implications of our findings for galaxy-size predictors commonly employed in semi-analytic and empirical models.
We summarize our conclusions in \se{conclusion}.
Throughout this work, we define haloes as spherical overdensities of 200 times the critical density of the Universe at $z=0$, with $\rhoc=126.08 \Msun \kpc^{-3}$, and denote halo virial radius and mass as $\Rv$ and $\Mv$, respectively.

\section{Methods}\label{sec:method}

In this section, we present the details of our controlled simulation suite.
We describe and motivate the choices of halo parameters and the numerical configurations adopted in our models  (\se{setup}), summarize the numerical code \GIZMO and the \FIRE galaxy-formation prescriptions employed in our simulations (\se{code}), and outline the analysis tools used in this work (\se{analysis}). 

\subsection{Halo parameters and simulation setups} \label{sec:setup}

Here we focus on the impact of halo structural parameters on galaxy size and keep the halo mass fixed at $10^{11}\Msun$. 
We choose this mass scale for several reasons. 
First, structural diversity of galaxies is expected to be maximal at this mass \citep[e.g.,][]{Cintio14,Oman15,Lazar24}. 
Second, this mass lies near the low-mass side of the morphology transition identified in cosmological simulations: lower-mass galaxies are often kinematically hot or irregular, whereas well-developed rotationally supported discs become most common at somewhat higher galaxy and halo masses \citep[e.g.,][]{Clauwens18,Tacchella19,Dekel20,Liang25a,Benavides25}. 
Third, in the more massive Milky Way-scale haloes ($\sim10^{12}\Msun$), while discs are common, the influence of accreting supermassive black holes is, in principle, non-negligible. However, such effects are not yet fully captured by current subgrid prescriptions.

That said, we have verified with a set of test simulations spanning different halo masses that our models approximately reproduce key scaling relations, such as the stellar mass–halo mass relation, and exhibit the strongest cusp–core transitions in the halo mass regime characteristic of—and consistent with—those found in cosmological FIRE-2 simulations \citep{Lazar20}.

The parameters varied in our simulations are defined as follows.
\begin{itemize}[leftmargin=*]
\item \textbf{Halo concentration} is defined as $c_2=\Rv/r_{-2}$, where $r_{-2}$ is the radius at which the logarithmic density slope equals –2.
DM haloes are initialized and fitted using the Dekel-Zhao (DZ) profile \citep{Freundlich20}, which introduces one additional degree of freedom relative to the \citet[][NFW]{Navarro97} profile and allows direct control of the logarithmic density slope of the halo. 

The DZ profile is given by:
\be\label{eq:DZ}
\rho(r)=\frac{\rho_0}{x^a(1+x^{1/2})^{2(3.5-a)}},
\ee
where $x = r/\rc$, with $\rc$ a scale radius defining the concentration $c = \Rv/\rc$.
The parameter $a$ specifies the asymptotic inner density slope as $r\to 0$,
and $\rho_0$ is characteristic density related to the virial overdensity $\Delta$ via:
\be
\rho_0 = (1 - a/3)\overline{\rho_0},  \quad
\overline{\rho_0} = c^3 \mu \Delta \rhoc,
\ee
with 
\be
\mu = c^{a-3}(1 + c^{1/2})^{2(3-a)}.
\ee
The DZ parameters $c$ and $a$ are not always the most convenient for comparison with the literature. 
In particular, $c$ differs from the commonly used concentration parameter $c_2$, which coincides with the standard NFW definition. 
The two are related by
\be\label{eq:c2}
c_2\equiv\frac{\Rv}{r_2}=c\left(\frac{1.5}{2-a}\right)^2.
\ee

\item \textbf{Inner density slope}, $s_1$, is defined as the logarithmic slope $-\rmd \ln \rho / \rmd \ln r$ evaluated at 1\% of the virial radius.
This quantity can be expressed in terms of the DZ parameters $c$ and $a$ as:
\be\label{eq:s1}
s_1=\frac{a+3.5c^{1/2}(r_1/R_{\rm{vir}})^{1/2}}{1+c^{1/2}(r_1/R_{\rm{vir}})^{1/2}}=\frac{a+0.35c^{1/2}}{1+0.1c^{1/2}},
\ee
where $r_1=0.01\Rv$.

\item \textbf{Halo spin parameter} is defined following \citep{Bullock01b} as
\be\label{eq:spin}
\lambda=\frac{\jv}{\sqrt{2}\Rv\Vv},
\ee 
where $\jv=\Jv/\Mv$ is the specific angular momentum of the halo, and $\Vv$ the circular velocity evaluated at the virial radius $\Rv$.

\item \textbf{The baryon fraction} of a halo, $\fb$, is defined as the ratio of the total baryonic mass to the virial mass. 
Here, we emphasize that $\fb$ is not necessarily equal to the cosmic baryon fraction of $\simeq 0.165$.
Observations of nearby galaxies reveal substantial scatter in baryon fraction across a wide range of halo masses \citep[e.g.][]{ManceraPina25}. 
Consistent with this, cosmological hydrodynamical simulations such as FIRE-2 also show that $\fb$ deviates from the cosmic fraction at halo masses below the Milky Way  and exhibits significant scatter \citep{Hafen19}. 
We therefore include $\fb$ as a parameter in our grid and explore how variations in $\fb$ affect the resulting galaxy properties. 
We adopt a fiducial value of 0.07, motivated by the results of \citeauthor{Hafen19}
\end{itemize}

\begin{table}
  \centering
  \caption{Values adopted in the controlled simulations suite for each parameter in the factorial grid of $c_{2}$, $s_{1}$, $\lambda$, and $f_{\rm b}$, at a fixed halo mass of $\Mv=10^{11}\Msun$ (see Section \ref{sec:setup}). 
  The bold fonts indicate fiducial values.}
  \label{tab:Param}
  \begin{tabular}{lc}
    \toprule
    \textbf{Properties} & \textbf{Values} \\
    \midrule
    $c_{2}$                 & 3,\;\textbf{10},\;20 \\
    $s_{1}$                 & 0.0,\;0.5,\;\textbf{1.0},\;1.5 \\
    $\lambda$               & 0.02,\;\textbf{0.04},\;0.06,\;0.08 \\
    $f_{\rm b}$                 & 0.03,\;\textbf{0.07},\;0.14 \\
    \bottomrule
  \end{tabular}
\end{table}

We construct a nearly factorial parameter grid that samples the four halo properties  
($c_{2}$, $s_{1}$, $\lambda$, $f_{\rm b}$) at the discrete values listed in \Tab{Param}. 
These parameter choices are designed to sample the range of values commonly found in cosmological simulations. For example, halo concentration exhibits a lower bound of $\sim 3$, a median value of $\sim 10$, and a $\sim 2\sigma$ upper-tail value of $\sim 20$ at the halo mass of interest \citep[e.g.][]{DuttonMaccio14,DiemerKravtsov15}. 
Similarly, the halo spin parameter has a median value of $\simeq 0.04$, with 0.02 and 0.08 sampling the approximate $2\sigma$ tails of the distribution \citep{Bullock01b}. 
The inner density slope values span the full range relevant to the cusp–core problem \citep[e.g.][]{Bose19, Freundlich20}, from completely flat cores ($s_1 = 0$) to steep cusps ($s_1 = 1.5$).
In principle, this grid would yield $3\times4\times4\times3=144$ parameter combinations. 
However, the DZ functional form does not permit physically meaningful solutions for the combination $(c_2,s_1)=(20,0)$, and we therefore exclude all models with this pairing. 
The resulting design consists of $N=132$ distinct simulation runs, which form the main sample analyzed in this work.

For each parameter combination, we construct an initial condition (IC) consisting of a DZ DM halo and a gaseous component in approximate hydrostatic equilibrium. 
There is no unique prescription for establishing exact  equilibrium between the gas and the host halo; instead, we follow the approach of \citet{KimLee13}, in which the gas distribution that is initialized to be self-similar to the DM halo and is then allowed to relax toward equilibrium under gravity and hydrodynamics alone.

In practice, we first generate two collisionless haloes with identical structural parameters but particle numbers different by a factor of five, using $10^7$ and $2\times10^6$ particles, respectively, with the MAny-component Galaxy Initializer (\MAGI) code. 
This corresponds to a gas particle mass of $10^4 M_\odot$ and a dark-matter particle mass of $5\times10^4 M_\odot$, with a minimum adaptive gravitational softening of 0.5 pc for gas and constant softening of 30 pc for dark matter, respectively, comparable to the numerical choices by \citep{Hopkins18}.
We then combine these two realizations by randomly designating a fraction $\fb$ of the higher-resolution particles as gas particles and a fraction $(1-\fb)$ of the lower-resolution particles as DM particles.
The resulting composite system is subsequently relaxed in the hydrodynamics solver \GIZMO \citep{Hopkins15} under gravity and hydrodynamics only. 

The initial temperature of the gaseous component is set to the halo virial temperature,
\be
T_{\rm{vir}}=\frac{\mu m_{\rm p}}{2k_\mathrm{B}}\frac{G\Mv}{\Rv}.
\ee
We adopt a uniform initial gas metallicity of $Z=0.003Z_{\odot}$, following \citet{KimLee13}. 

To set the halo spin, we adopt the empirical specific angular-momentum profile proposed by \citet{Bullock01b},
\be
j(<M) = g_0 M^{1.3}.
\ee
Where $j(<M)$ denotes the cumulative specific angular momentum enclosed within mass $M$. 
We determine the normalization $g_0$ by requiring that the integrated angular momentum of the halo reproduces a target spin parameter\footnote{This trial target value is different from (slightly higher than) the intended grid value listed in \Tab{Param}, because the full-physics simulations are initialized after the system has relaxed, and all quoted values correspond to measurements taken after this relaxation phase.}. 
Given $g_0$, we assign the corresponding specific angular momentum to each mass shell and adjust particle velocities accordingly.

We then evolve this configuration for 3 Gyr under gravity and hydrodynamics only, allowing the system to relax. 
During this relaxation phase, the properties of the gaseous halo evolve slightly. 
The halo parameters listed in our experimental design (\Tab{Param}) therefore correspond to the values measured after this relaxation period. 
The relaxed systems are subsequently evolved for an additional 3 Gyr using the \GIZMO code with full physics enabled. 
This duration is slightly longer than the dynamical time of a $z=0$ halo and is sufficient for the galaxies to reach a quasi-steady state.
We analyze the galaxy properties at the end of the full-physics simulations and quote the halo properties at the conclusion of the relaxation phase. 

\subsection{Overview of the GIZMO code and the FIRE-3 physics}\label{sec:code}

The simulations are performed using \GIZMO \citep{Hopkins15}, a multi-method gravity plus (magneto)hydrodynamics code, operated in its mesh-free finite-mass (MFM) mode.  
Gravitational forces are computed using an improved Tree-PM solver from {\tt GADGET-3} \citep{Springel05a}.

The simulations incorporate the \FIRE physics model \citep{Hopkins23}, which builds on the \texttt{FIRE-2} framework with updated stellar evolution models, metal yields, and microphysical prescriptions for gas cooling, heating, and feedback. 
Gas cooling and heating processes are modeled over a wide temperature range, from $10$ to $10^{10}$ K, including approximate treatments of self-shielding of dense gas from the UV background and from local radiation sources via the LEBRON scheme \citep{Hopkins20}. 
Metal-line cooling follows \citet{Wiersma09} at temperatures $> 10^{4}$ K, and pre-tabulated {\tt CLOUDY} rates \citep{Ferland98} are used at lower temperatures.

Star formation occurs in gas that is locally self-gravitating \citep{Hopkins13}, in a converging flow, Jeans unstable, and denser than $n_{\rm th}=1000\,{\rm cm}^{-3}$. 
Once these conditions are met, stars form with 100\% efficiency per local free-fall time. 
Each star particle represents a single stellar population, assuming a \citet{Kroupa01} initial mass function for stars over the mass range $0.1-100\,\Msun$.

Stellar feedback includes momentum input from radiation pressure in the UV, optical, and infrared, with radiative-transfer effects approximated using the LEBRON scheme, as well as energy, momentum, mass, and metal injection from Type Ia and core-collapse supernovae and stellar mass loss \citep{Fielding18}. 
Local photoionization and photoelectric heating from young stars are also modeled. 

\subsection{Simulation analysis}\label{sec:analysis}

We developed a custom analysis pipeline for this simulation suite, built primarily on \texttt{PYNBODY} \citep{Pontzen13}, while incorporating selected routines and conventions from \texttt{GizmoAnalysis} \citep{Wetzel20} and \texttt{YT} \citep{Turk11}.
For each snapshot, we re-centre the system using the \texttt{PYNBODY} centring routines and rotate the coordinate frame so that the z axis aligns with the direction of the total stellar angular momentum. 
We then measure halo and galaxy properties.
To characterize the host-halo structure, we fit the spherically averaged DM density profile with the DZ profile to obtain the structural parameters $c_2$ and $s_1$ using \eq{c2} and (\ref{eq:s1}). 
We use the three-dimensional stellar half-mass radius, $r_{1/2,\star}$, as the main definition for galaxy size. 
\footnote{We opt for this morphology-independent metric instead of a disk scale radius because galaxies at this mass scale do not always form stable discs. The simple half-mass radius allows us to robustly characterize the overall spatial extent of the system regardless of whether it is discy, spheroidal, or irregular.}
We also examine the half-mass radius of stars and cold gas with temperature $T<10^{4}$ K, $r_{1/2,\star+{\rm cg}}$, which we refer to as the size of cold baryons. 

To quantitatively characterize the impact of halo parameters on galaxy size, we fit a second-order (quadratic) response-surface model (RSM; \citealt{Myers2016RSM}) to $y\equiv\log r_{1/2,\star}$ as a function of the four halo parameters.
We adopt the predictors $\log c_2$, $\log \lambda$, $\log f_{\rm b}$, and $s_1$, and standardize each predictor $u_k$ using its midpoint $u_{k,0}$ and half-range $\Delta u_k$ via $z_k = (u_k-u_{k,0})/\Delta u_k$, where $z_k\in[-1,1]$. 
The response surface is then modeled as
\be\label{eq:rsm model}
    y \approx \beta_0 
      + \sum_{k} \beta_{1,k}\,z_k
      + \sum_{k} \beta_{2,k}\,z_k^2
      + \sum_{i<j} \beta_{ij}\,z_i z_j,
\ee
where $\beta_0$ is the intercept, the coefficients $\beta_{1,k}$ represent linear terms, $\beta_{2,k}$ quantify curvature, and $\beta_{ij}$ capture pairwise interaction terms.  
With standardized predictors, $\beta_{1,k}$ and $\beta_{2,k}$ describe the local sensitivity and curvature of the response around the fiducial model, respectively, while $\beta_{ij}$ quantify non-additive couplings between parameters.
RSM provides a smooth approximation to the galaxy size across the parameter space and enables an explicit decomposition into main effects (linear plus quadratic terms) and pairwise interactions. 
Additional details are provided in Appendix~\ref{app:rsm}.

\begin{figure*}
    \centering
    \includegraphics[width=\textwidth]{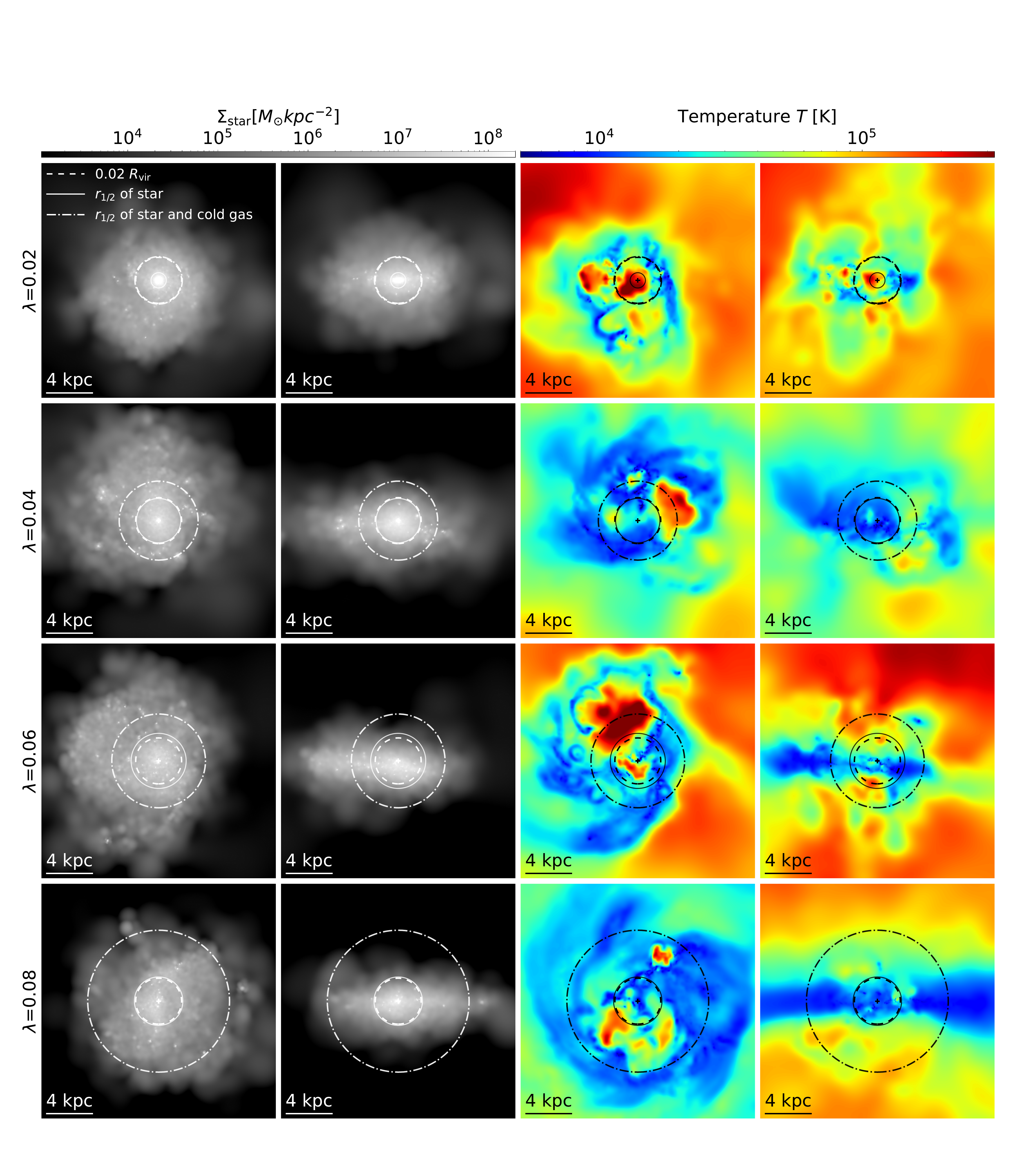}
    \caption{{\bf Projected stellar surface density and gas temperature maps} for simulations in which the halo spin parameter $\lambda$ is varied from 0.02 to 0.08 (one value per row, as indicated), while all other parameters are held at their fiducial values: concentration $c_2=10$, logarithmic dark-matter-density slope $s_1=1$, and baryon fraction $\fb=0.07$. 
    Columns 1 (2) and 3 (4) show the face-on (edge-on) views. 
    Circles indicate $0.02\,R_{\rm vir}$ (dashed), the stellar half-mass radius $r_{1/2,\star}$ (solid), and the half-mass radius of stars plus cold gas $r_{1/2,\star+{\rm cg}}$ (dash-dotted).
    }
    \label{fig:maps_spin}
\end{figure*}

\begin{figure*}
    \centering
    \includegraphics[width=\textwidth]{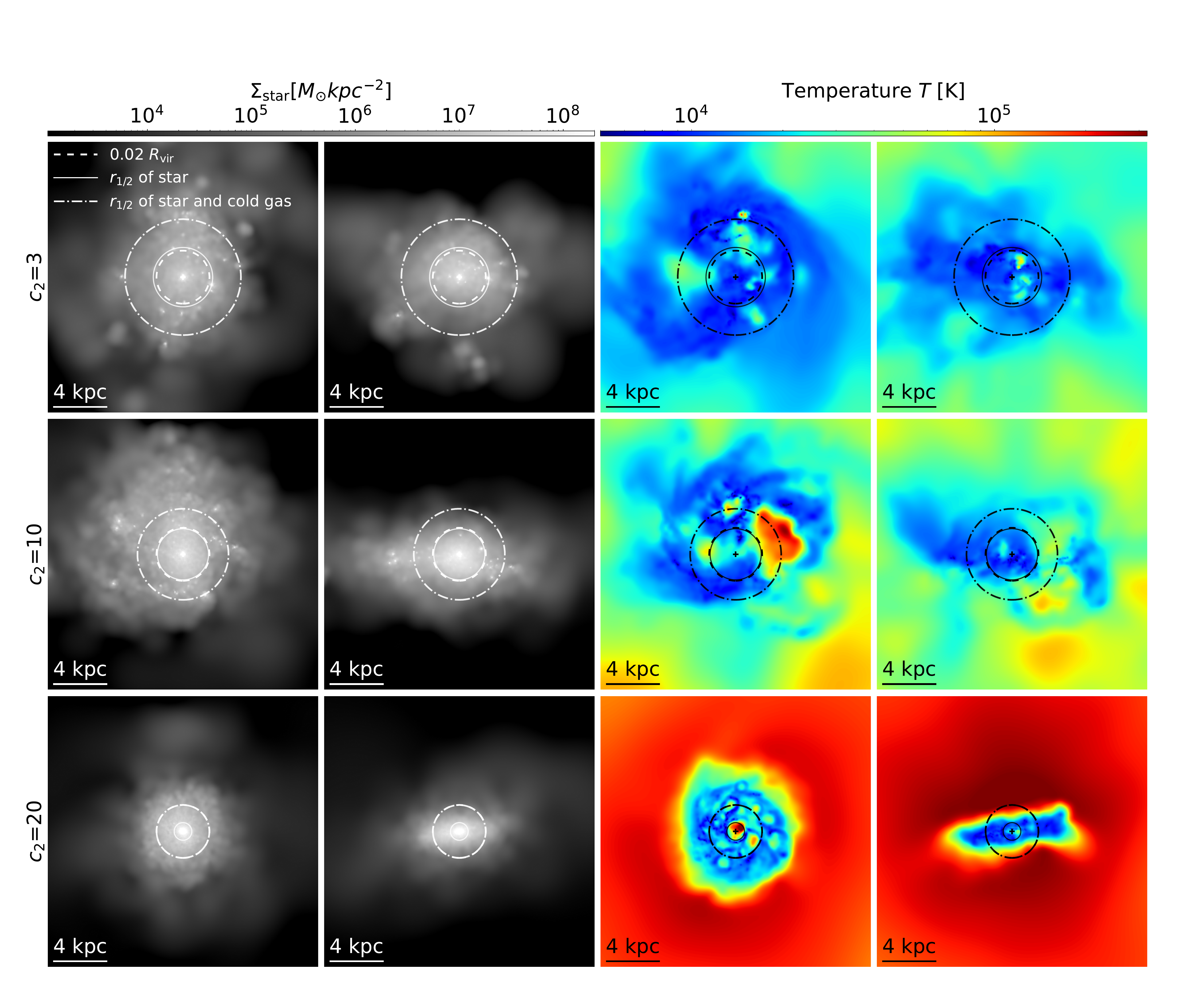}
    \caption{Same as Fig.\ref{fig:maps_spin}, but varying halo concentration $c_2$ from 3 to 20 while keeping all other parameters at their fiducial values: spin $\lambda=0.04$, inner DM density slope $s_1=1$, and baryon fraction $\fb=0.07$.}
    \label{fig:maps_c2}
\end{figure*}

\begin{figure*}
    \centering
    \includegraphics[width=\textwidth]{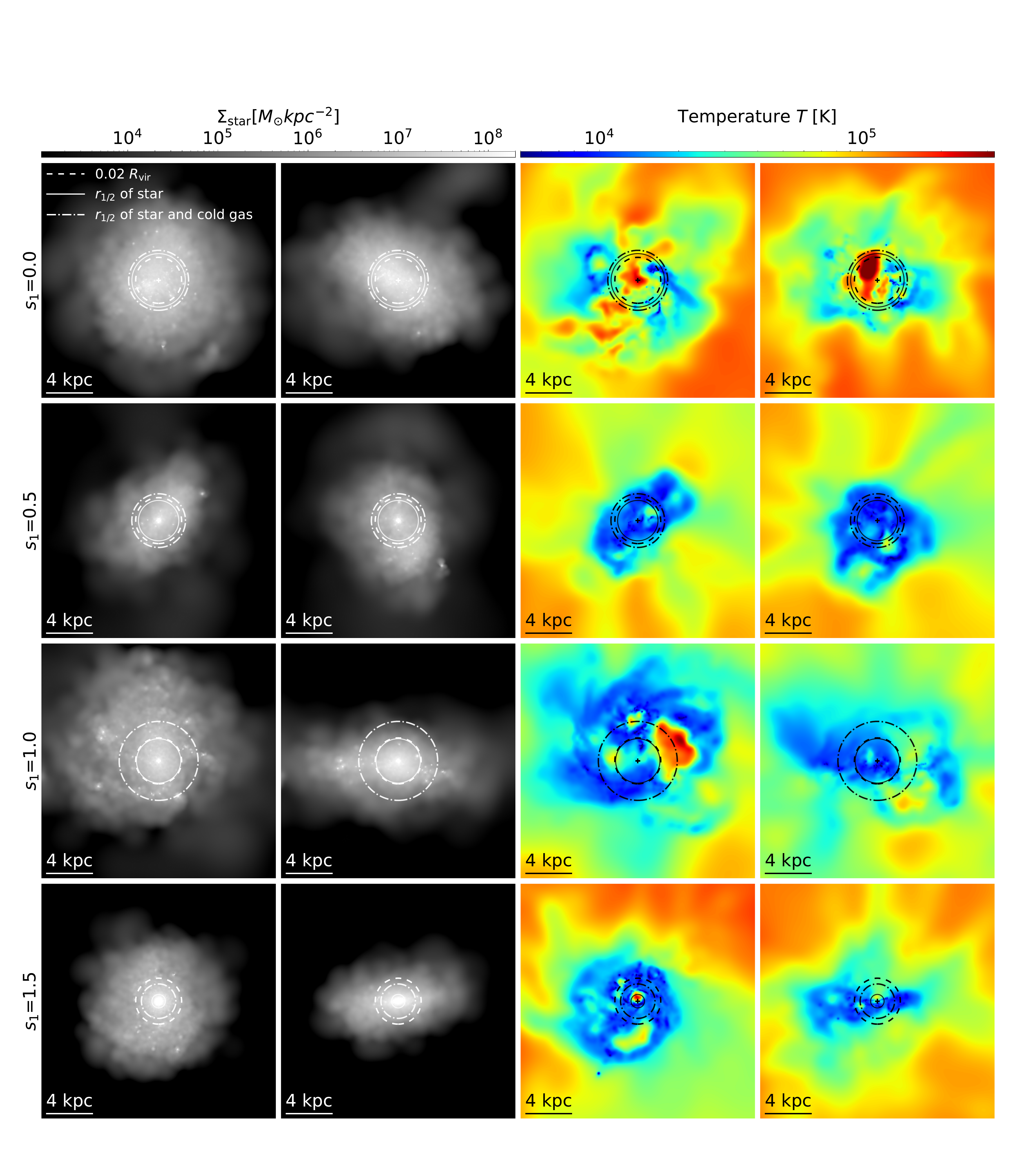}
    \caption{Same as Fig.~\ref{fig:maps_spin}, but varying the inner logarithmic slope of dark matter density $s_1$ from 0 to 1.5, while keeping all other parameters at their fiducial values: spin $\lambda=0.04$, concentration $c_2=10$, and baryon fraction $\fb=0.07$.}
    \label{fig:maps_s1}
\end{figure*}

\begin{figure*}
    \centering
    \includegraphics[width=\textwidth]{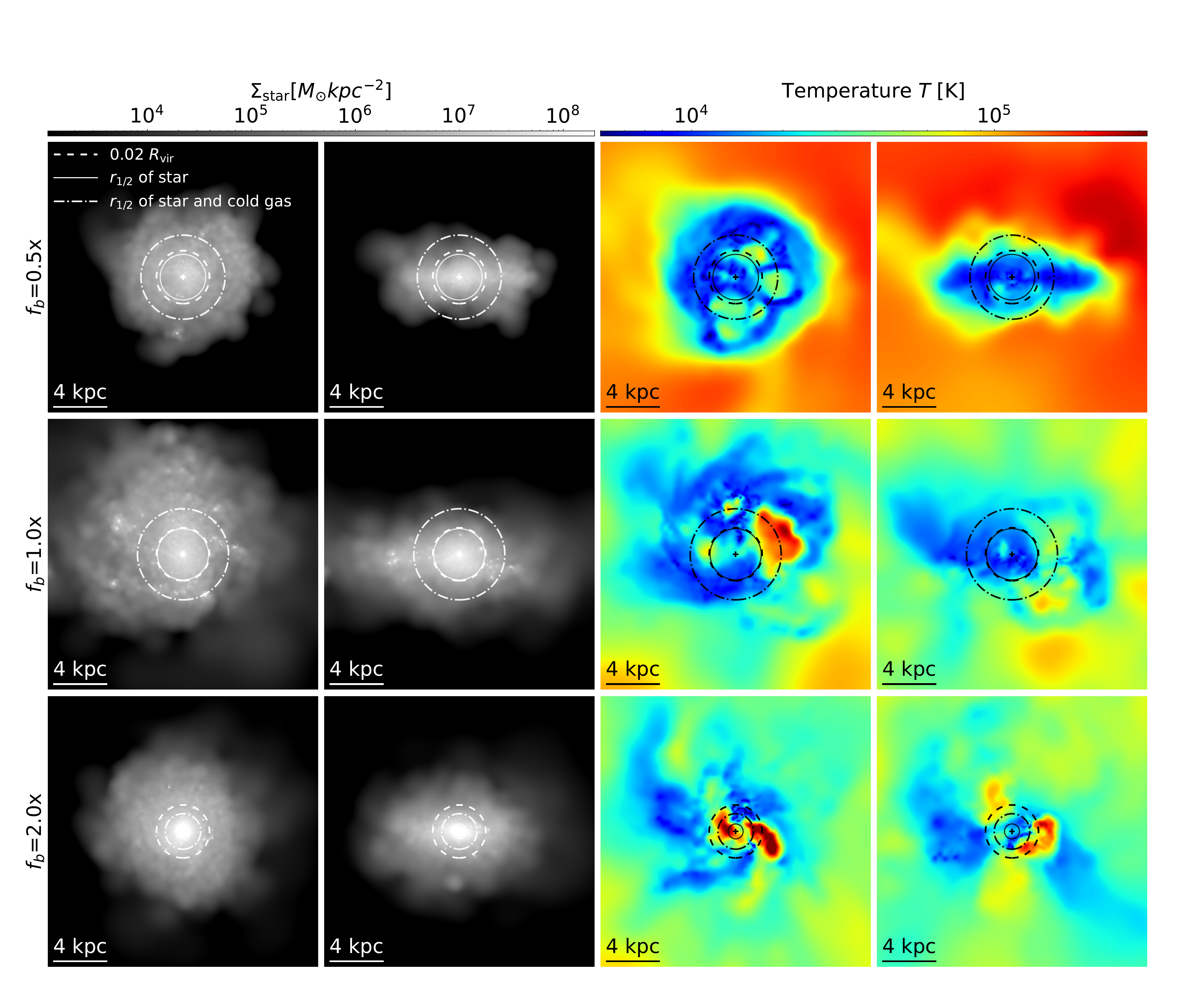}
    \caption{Same as Fig.~\ref{fig:maps_spin}, but varying the baryon fraction $f_{\rm b}$ relative to the fiducial value of 0.07 \citep{Hafen19} from $0.5\times$ to $2.0\times$, while keeping all other parameters at their fiducial values: spin $\lambda=0.04$, concentration $c_2=10$, and logarithmic dark-matter-density slope $s_1=1$.}
    \label{fig:maps_fb}
\end{figure*}

\section{Results}\label{sec:results}

In this section, we present results from our controlled simulation suite to examine how galaxy size responds to systematic variations in halo structural parameters and baryon fraction at fixed halo mass. 
We begin with a qualitative overview based on projected density and gas temperature distributions (\se{morphology}). 
We then quantify these trends across the full parameter grid using a quadratic response-surface method to identify the dominant effects (\se{rsm}). As a complementary approach, we also employ a random-forest regressor to assess the relative importance of the halo parameters regulating galaxy size (\se{random_forest}).

\subsection{Galaxy size under variation of halo parameters}
\label{sec:morphology}

\Figs{maps_spin}-\ref{fig:maps_fb} present the stellar surface density distributions as well as gas temperature maps for a set of simulations where only one halo parameter is varied while the others are fixed at the fiducial values (\Tab{Param}). 
Overlaid on these maps are three reference radii: 2 percent of the virial radius, $0.02 R_{\rm vir}$, the half-stellar-mass radius $r_{1/2,\star}$, and the half-mass radius of stars and cold gas $r_{1/2,\star+{\rm cg}}$.

In general, our simulations yield a median $r_{1/2,\star}$ of 1.65 kpc with the 16th and 84th percentiles being 0.63 kpc and 3.66 kpc respectively. 
Reassuringly, as a reference, the standard scaling relation of $r_{1/2,\star}=0.02R_{\rm{vir}}$ with a scatter of 0.2 dex \citep{Kravtsov13,Somerville18}, corresponds to a median of 1.92 kpc and a range of 1.21 to 3.04 kpc at this halo mass, comparable to the range spanned by our simulated galaxies. 
Hence, the systematic variations in halo structure explored here can generate the desired diversity of galaxy sizes. 
\footnote{We caution that our simulation suite is a uniform experimental design rather than a cosmological sample, so the percentiles should be interpreted as coverage of the explored parameter space rather than the intrinsic population scatter at fixed halo mass.}

Increasing the halo spin leads to a clear growth in galaxy size, as shown in \Fig{maps_spin}. 
This effect is particularly pronounced for the cold-baryon half-mass radius, $r_{1/2,\star+{\rm cg}}$, which increases by nearly a factor of three as the spin parameter increases from 0.02 to 0.08. In contrast, the stellar half-mass radius, $r_{1/2,\star}$, grows more moderately and appears to saturate at $\lambda \ga 0.06$. 
These trends are qualitatively consistent with the theoretical expectations of \citet{Mo98}; however, they already indicate that galaxy size does not scale linearly with halo spin. 
We revisit this point in \se{discussion}.

Increasing the halo concentration produces an opposite trend, with higher $c_2$ yielding systematically smaller discs, as shown in \Fig{maps_c2}. This effect is generally more pronounced for the cold-baryon half-mass radius, $r_{1/2,\star+{\rm cg}}$, than for the stellar half-mass radius, $r_{1/2,\star}$. 
As the cold disc becomes more compact, the hot gaseous halo reaches deeper into the gravitational potential, as can be seen from the temperature maps in the right columns of \Fig{maps_c2}.

The inner density slope of the DM halo has a more subtle influence on galaxy morphology. 
While steeper inner slopes tend to enhance the central gas density, the overall galaxy size scales remain similar for $s_1 \lesssim 1$, particularly for the stellar component, whose size remains close to $0.02\,R_{\rm vir}$. 
For the most cuspy case, $s_1 = 1.5$, however, galaxy sizes, especially the stellar half-mass radius, are significantly reduced.

Varying the baryon fraction $\fb$ has a significant impact on galaxy morphology and size. 
As shown in \Fig{maps_fb}, for $\fb \lesssim 0.07$, galaxy sizes are  similar, with stellar half-mass radii $r_{1/2,\star} \sim 0.02 R_{\rm vir}$ and cold-baryon half-mass radii $r_{1/2,\star+{\rm cg}}$ approximately twice as large. 
In this low-$\fb$ regime, the cold gas forms a relatively settled disc, as clearly indicated by the temperature maps.
In contrast, at $\fb = 0.14$, which is close to the cosmic baryon fraction, the system undergoes centrally concentrated star formation and develops a compact stellar component. 
Even the cold gas is no longer arranged in a well-ordered, disc-like configuration. 
These results imply that the baryon fraction strongly regulates where gas can reach star-formation conditions, thereby shaping both galaxy morphology and size -- an effect that becomes particularly pronounced at high $\fb$.

We caution that the $\fb$ dependence may be exaggerated in our idealized setup relative to fully cosmological simulations. 
In our high-$\fb$ model, an amount of gas that is close to the cosmic baryon fraction is initialized instantaneously within the moderately massive halo, rather than being accreted gradually over cosmic time as in a gradually growing halo. 
As a result, dense central gas may instantly satisfy the star-formation criteria, triggering a central starburst. 
Despite this caveat, our simplified experiments offer valuable physical insight.
First, they suggest that any episode of high baryon fraction in a moderately massive halo (e.g., $\Mv\sim10^{11}\Msun$) likely leads to the formation of a compact galaxy, even when strong stellar feedback is included through the \FIRE\ physics. 
Second, baryon fractions below the cosmic mean are a natural requirement for bright dwarf systems to develop extended sizes and discy morphologies.

\subsection{Quantifying the impact of halo properties on galaxy size}
\label{sec:rsm}

\begin{table}
  \centering
  \caption{
  Summary of complementary importance proxies used to quantify the dependence of the stellar half-mass radius, $r_{1/2,\star}$, on the four halo parameters. 
  Spearman's rank coefficient $r_{\rm sp}$ measures the monotonic correlation between each parameter and $r_{1/2,\star}$ across the full simulation sample, with its sign indicating the direction of the trend.
  The quadratic response-surface model (RSM) provides two conditional effect-size metrics: a main-effect partial $R^2$, obtained from a joint (linear + quadratic) coefficient-block test for each parameter, and a total-effect partial $R^2$, derived from an analogous block test that additionally includes all pairwise interaction terms involving that parameter (definitions and HC3-robust inference are described in Appendix~\ref{app:rsm}). 
  Random-forest (RF) entries report normalized feature importance (each RF row sums to unity across the four parameters); we list four RF estimators—MDI, out-of-fold permutation, OOB permutation, and SHAP (Appendix~\ref{app:rf}). 
  }
  \label{tab:importance}
  \begin{tabular}{lcccc}
    \toprule
    \textbf{importance proxy} & \textbf{$c_{2}$} & \textbf{$\lambda$} & \textbf{$s_{1}$} & \textbf{$f_{\rm b}$}\\
    \midrule
    Spearman $r_{\rm sp}$              & $-0.57$ & $+0.44$ & $-0.51$ & $+0.07$ \\
    \midrule
    RSM main-effect partial $R^2$     & 0.463   & 0.477   & 0.492   & 0.033   \\
    RSM total-effect partial $R^2$    & 0.579   & 0.523   & 0.544   & 0.210   \\
    \midrule
    RF MDI                & 0.303   & 0.241   & 0.267   & 0.188   \\
    RF permutation (OOF)    & 0.389   & 0.231   & 0.200   & 0.180   \\
    RF permutation (OOB)    & 0.395   & 0.235   & 0.193   & 0.178   \\
    RF SHAP                & 0.372   & 0.245   & 0.255   & 0.128   \\
    \bottomrule
  \end{tabular}
\end{table}

The visual examinations of the density maps and temperature maps in \se{morphology} confirm that all the four parameters impact galaxy morphology and size.
It is natural to aim for a quantitative result on the relative importance. 
Since there is no simple concordance metric for relative importance, we adopt complementary measurements summarized in Table~\ref{tab:importance}. 

In this section, we consider two importance proxies. 
The first is the Spearman rank correlation coefficient $r_{\rm sp}$ between each halo parameter and galaxy size $r_{1/2,\star}$. 
Larger $|r_{\rm sp}|$ indicates a stronger monotonic trend.  
The second class of proxies is derived from the fitted quadratic response-surface model (RSM; Section~\ref{sec:analysis}, eq.~\ref{eq:rsm model}).
For a given factor, we assess its contribution by comparing the full RSM to a  reduced model in which selected terms involving that factor are removed and the remaining coefficients are re-estimated. 
The resulting degradation in fit defines a partial $R^2$ (coefficient of partial determination), namely the fraction of the residual variance under the reduced model that is explained by the removed terms. 
We report both a \emph{main-effect} partial $R^2$, obtained by removing the linear and quadratic terms of that factor, and a \emph{total-effect} partial $R^2$, which additionally removes all interaction terms involving the same factor (see Appendix~\ref{app:rsm} for details).

\Fig{rsm_main_effect} shows the correlations between the stellar half-mass radii and individual halo parameters in our simulation suite. 
Despite substantial scatter at fixed parameter values, the mean trends are clear and broadly consistent with the visual impressions from Figs.~\ref{fig:maps_spin}–\ref{fig:maps_fb}.
In particular, $r_{1/2,\star}$ increases with halo spin $\lambda$, with a correlation coefficient of $r_{\rm sp}=0.44$, and decreases with both concentration $c_2$ and inner density slope $s_1$, with $r_{\rm sp}=-0.57$ and $r_{\rm sp}=-0.51$, respectively. 
The overall rank correlation with $f_{\rm b}$ is nearly null ($r_{\rm sp}=0.07$), seemingly at odds with the visual impression from \Fig{maps_fb}. 
We caution, however, that this is mainly because the dependence on $\fb$ is often non-monotonic at fixed halo structural parameters.
This behavior is evident from the thin lines connecting simulations in the upper right panel of \Fig{rsm_main_effect}: several runs exhibit a mild increase in galaxy size with $\fb$ at $\fb\lesssim 0.07$, followed by a decrease at larger $\fb$, a trend that is particularly pronounced for some cases with $c_2=3$ and $10$.
We have verified that the non-monotonic behavior of the $\fb$ trend is also present at moderate halo spin values ($\lambda\simeq 0.04$-0.06). 
Only at the highest spin, galaxy size increases monotonically with $\fb$, whereas at the lowest spin values galaxy size decreases monotonically with $\fb$.

\begin{figure*}
    \centering
    \includegraphics[width=0.7\textwidth]{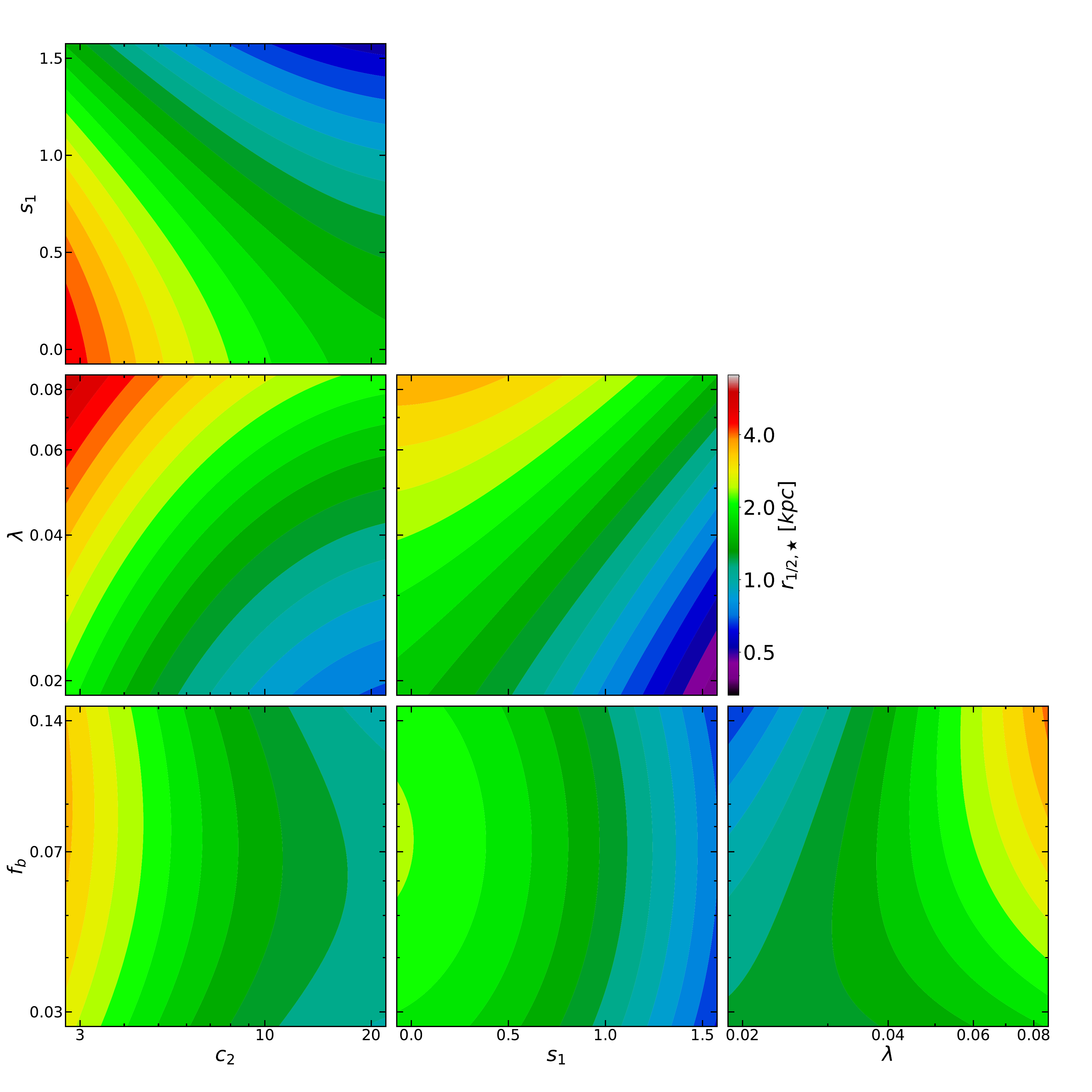}
    \caption{Galaxy size (half-stellar mass radius) as a function of halo structural parameters and baryon fraction, as predicted by the response-surface model fitted to our simulation results. 
    In each panel spanning a pair of parameters, the remaining two parameters are held fixed at their midpoint values. 
    The halo structural parameters exhibit largely additive behavior, reflected by quasi-parallel contours. In contrast, panels involving the baryon fraction show non-separable responses, with the $\fb$–$\lambda$ plane displaying the strongest covariance. This indicates that the baryon fraction modulates the sensitivity of galaxy size to variations in halo structure.
    }
    \label{fig:rsm_interaction}
\end{figure*}

This behavior is illustrated more clearly in Fig.~\ref{fig:rsm_interaction}, which visualizes the best-fit response-surface model in two dimensions by varying pairs of halo parameters at a time. 
The parameter spaces spanned by the halo structural parameters $c_2$, $s_1$, and $\lambda$ exhibit nearly linear gradients, indicating that the effects of individual parameters can, to a good approximation, be treated as additive. 
A notable exception arises in the parameter spaces involving the baryon fraction $\fb$, where the contours display pronounced curvature.
This behavior can be interpreted in two complementary ways. 
First, as noted above, at fixed $c_2$, $s_1$, and $\lambda$, the dependence of galaxy size on $\fb$ is generally non-monotonic. 
Second, the influence of halo concentration and spin on galaxy size becomes significantly stronger at higher baryon fractions ($\fb \gtrsim 0.1$).  
In this sense, the baryon fraction primarily modulates how sensitively galaxy size responds to halo structure, rather than acting as an independent regulator of size.

\begin{figure*}
    \centering
    \includegraphics[width=0.7\textwidth]{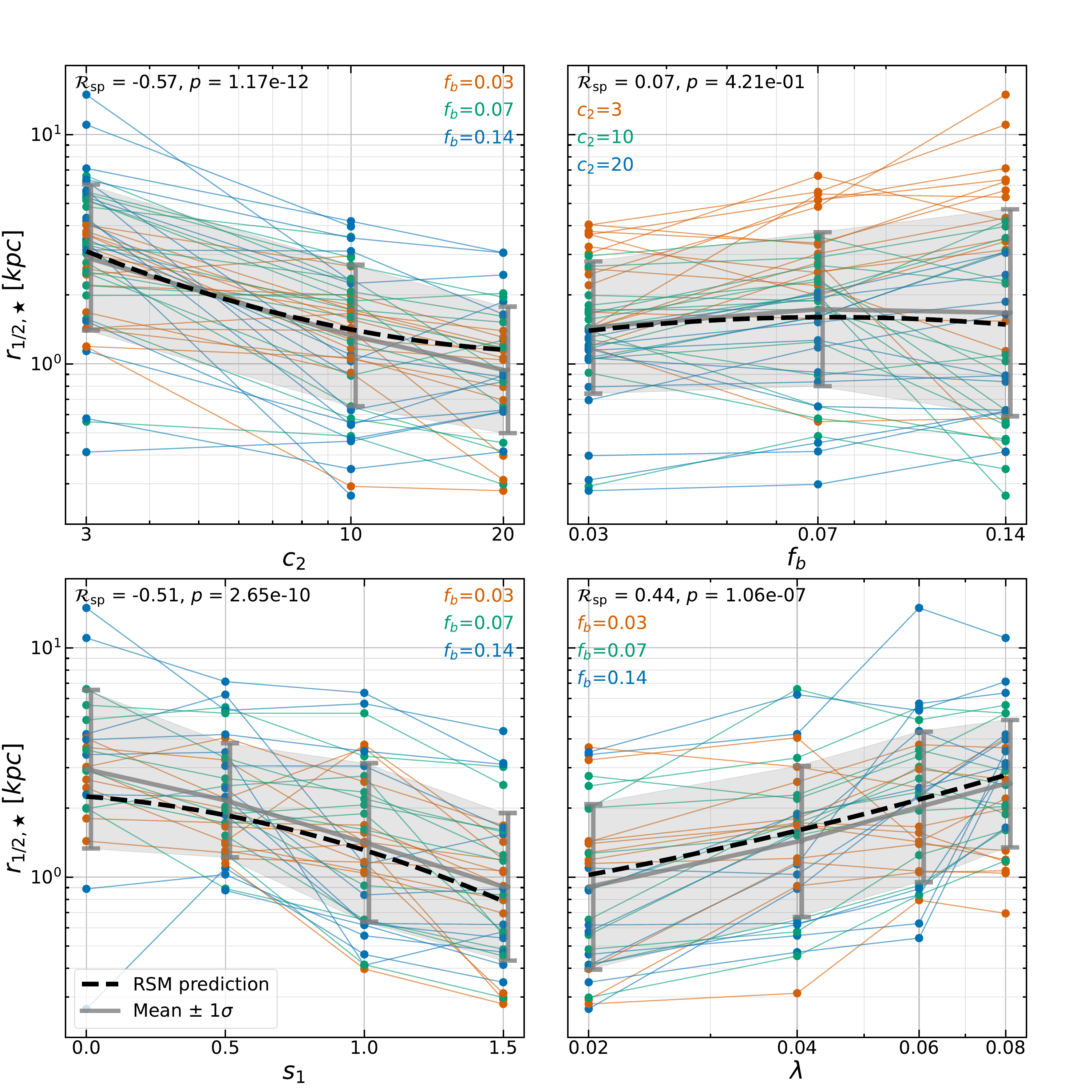}
    \caption{
    Galaxy size (half-stellar mass radius) as a function of halo structural parameters and baryon fraction. 
    Points represent individual simulation results, coloured by a secondary parameter ($\fb$ or $c_2$, as indicated). 
    Thin, colour-matched lines connect simulations in which only the horizontal-axis quantity is varied while all other parameters are fixed.  
    Connected grey symbols show the mean stellar half-mass radius, $r_{1/2,\star}$, with error bars and the light-grey bands indicating the $1\sigma$ scatter. 
    The symbol positions are slightly offset horizontally to improve visual clarity
    The black dashed curve shows the mean prediction of the best-fit quadratic response-surface model (RSM),  obtained by varying the parameter of interest while fixing all others at their midpoint values. 
    Spearman's rank correlation coefficients, $r_{\rm sp}$, and the corresponding $p$-values are quoted.
    }
    \label{fig:rsm_main_effect}
\end{figure*}

\subsection{Feature importance ranking from random-forest regression}
\label{sec:random_forest}

As a non-parametric complement to the quadratic response-surface model, we train a random-forest regressor \citep{Breiman01} to predict the stellar half-mass radius $r_{1/2,\star}$ from the four halo parameters $(c_2, s_1, \lambda, f_{\rm b})$ with {\tt sklearn} \citep{scikit-learn}.
Here our goal is not an optimal predictive model, but an importance ranking that is insensitive to the quadratic functional assumption in Section~\ref{sec:rsm}.
After experimenting with multiple importance estimators (Appendix~\ref{app:rf}), the same qualitative hierarchy emerges: halo concentration $c_2$ is consistently the most informative predictor of galaxy size, with a relative importance score of $\sim0.3$-0.4 depending on the specific metric, while halo spin $\lambda$ and inner slope $s_1$ have comparable contributions of $\sim$0.2-0.25, followed by the baryon fraction $f_{\rm b}$, which scores $\sim$0.15. 

Table~\ref{tab:importance} summarizes the importance metrics and scores discussed throughout this section. 
Halo concentration stands out as the primary size regulator followed by the spin parameter.  

\section{Discussion}\label{sec:discussion}

In this section, we compare our simulation results with existing models that predict galaxy size from halo properties and examine how different definitions of galaxy size affect the performance of these size predictors.

\subsection{Galaxy size predictors revisited}\label{sec:SizePredictors}

\subsubsection{Comparison with the Mo98 model}\label{sec:MMW98}

The classical disc-formation model of \citet{Mo98} (Mo98) predicts that galaxy size is primarily determined by the virial radius and spin of the host halo, under the assumption that the 
baryons forming the galaxy initially share the same specific angular momentum as the halo and retain a fraction of it during galaxy formation. 
In this framework, the disc half-mass radius is given by
\be\label{eq:Mo98}
r_{\rm 1/2,\star} = \frac{1.678}{\sqrt{2}} f_j \lambda \Rv f_c^{-1/2} f_R ,
\ee
where the size scales linearly with the {\it angular-momentum retention factor},
\be
f_j \equiv j_\rmd/m_\rmd.
\ee 
Here $m_\rmd=\Ms/\Mv$ and $j_\rmd=J_\rmd/\Jv$ denote the mass and angular-momentum fractions of the galaxy relative to its host halo, respectively.
The factor $f_c$ characterizes the energy structure of the halo and is defined as the ratio of the halo's binding energy to that of a singular isothermal sphere of the same mass.
The factor $f_R$ accounts for the contraction of the halo in response to the gravitational potential of the disc. 

The original Mo98 formula assumes that haloes follow an NFW profile.
Here, our simulations adopt the DZ profile in order to allow explicit control over the inner density slope. 
To enable a fair comparison, we therefore adapt the Mo98 model to DZ haloes and compute a predicted stellar half-mass radius $r_{1/2,\star}^{\rm (Mo98)}$ for each simulation. 
In practice, this entails computing the factors $f_c$ and $f_R$ factors using the DZ profile, and measuring $j_\rmd$ and $m_\rmd$ directly from the simulated galaxies, strictly following their original definitions based on the stellar mass and stellar angular momentum. 
Details of our implementation of the Mo98 model for DZ haloes, as well as validation tests, are provided in Appendix~\ref{app:mmw_dz}.

The left panel of \Fig{predictor_compare} compares the Mo98-predicted sizes, $r_{1/2,\star}^{\rm (Mo98)}$, with the stellar half-mass radii measured in the simulations, $r_{1/2,\star}$. 
The model predictions exhibit a clear systematic offset relative to the simulation results, in that the Mo98 model typically under-predicts galaxy sizes by a factor of $\sim 5$.
Aside from this systematic offset, however, the model reproduces an approximately linear scaling with the simulated galaxy sizes.

Colour-coding individual galaxies by the angular-momentum retention factor, $f_j$, reveals that the simulations typically yield values well below unity, spanning a wide range from $\sim 0.01$ to $\lesssim 1$, with a typical value around $f_j \sim 0.2$. 
To illustrate this point, we recompute the Mo98 size predictions after artificially boosting the retention factor by a factor of five, as indicated by the open triangles in \Fig{predictor_compare}. 
This ad hoc correction brings the predicted sizes into close agreement with the one-to-one relation.
The failure of the simple angular-momentum retention assumption is expected if star-forming gas is preferentially drawn from the low-angular-momentum tail of the gas distribution and if stellar feedback redistributes angular momentum. 

\begin{figure*}
    \centering
    \includegraphics[width=\textwidth]{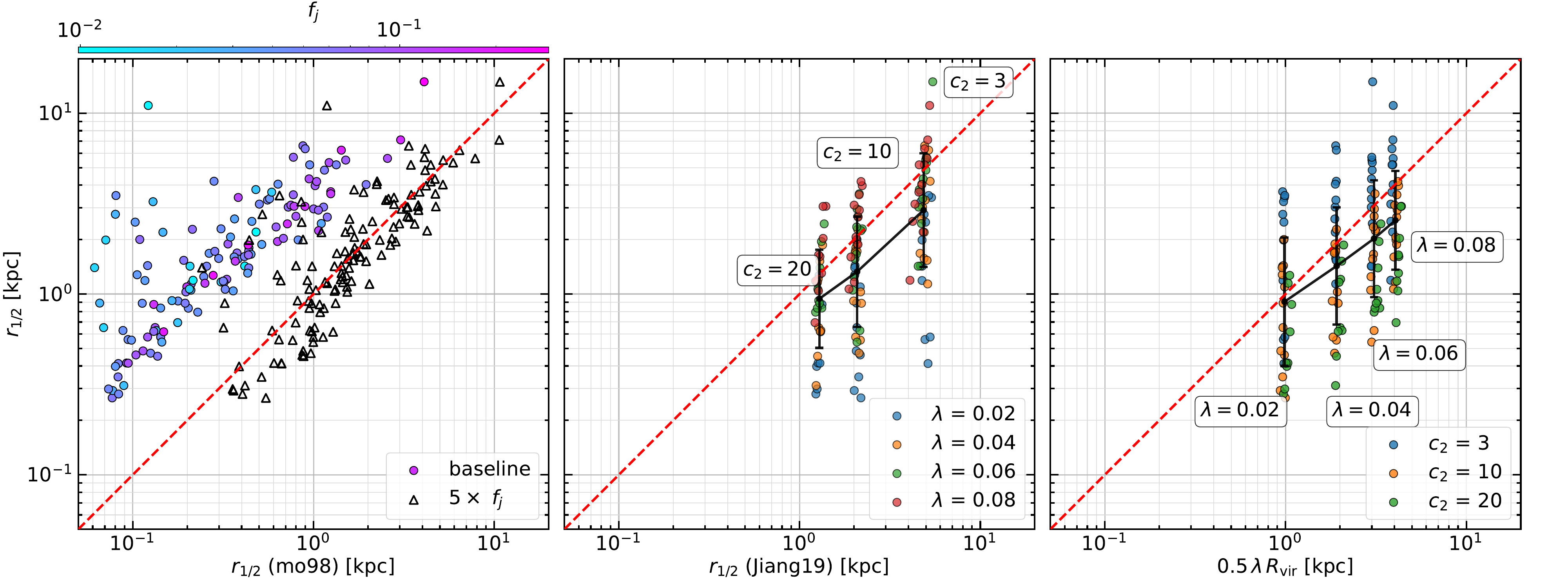}
    \caption{
    Comparison of simulated galaxy sizes with different size predictors. 
    In all panels, the y-axis shows the measured stellar half-mass radius, $r_{1/2,\star}$, and the red dashed line marks equality
    \quad
    \emph{Left panel:} the \citet{Mo98} (Mo98) predictor. 
    The x-axis shows the predicted stellar half-mass radius $r_{1/2,\star}^{\rm (Mo98)}$. 
    Filled circles show the predictions and are coloured by the angular-momentum retention factor $f_j\equiv j_{\rm d}/m_{\rm d}$. 
    Open triangles illustrate the effect of boosting $f_j$ by a factor of 5, which largely removes the systematic underprediction.
    \quad
    \emph{Middle panel:} the \citet{Jiang19} (J19) concentration-corrected predictor. 
    The x-axis shows the size predicted from the J19 relation $r_{1/2,\star}\simeq 0.02(c_2/10)^{-0.7}\Rv$. 
    The three discrete bands of dots correspond to the three  $c_2$ values in our design, as annotated. 
    Symbol colours indicate halo spin $\lambda$. 
    \quad
    \emph{Right panel:} a simple predictor based on the spin parameter: $r_{1/2,\star}\simeq 0.5\lambda \Rv$.
    The discrete four narrow bands of dots correspond to the four $\lambda$ values used in our simulations, as annotated.
    Symbol colours indicate halo concentration $c_2$.
    \quad
    For both {\it Middle} and {\it Right} panels, the black points indicate the average $r_{1/2,\star}$ in each bin, with error bars representing the $1\sigma$ scatter. 
    While these simple predictors capture the overall trends, they leave substantial scatter at fixed concentration or spin, indicating that additional parameters are required to explain the full diversity.
    }
    \label{fig:predictor_compare}
\end{figure*}

\subsubsection{Comparison with simple empirical size predictors}\label{sec:Jiang19}

\citet{Jiang19} (J19) found that galaxy spin (and hence size) is only weakly correlated with halo spin in their cosmological zoom-in simulations. 
They proposed an empirical, concentration-based size predictor of the form 
\be\label{eq:Jiang19}
r_{1/2,\star}\simeq 0.02(1+z)^{-0.2}(c_2/10)^{-0.7}\Rv.
\ee
In this relation, the dependence on halo concentration was introduced in part to account for the strong redshift-dependence of the ratio between galaxy size and halo virial radius seen in their cosmological simulations -- because the concentration itself evolves with mass and redshift and naturally introduces an implicit redshift dependence. 
J19 noted that the Mo98 formalism also contains a concentration dependence via the factor $f_c$.
But this dependence is much weaker.
Moreover, the redshift dependence of the concentration alone cannot fully explain the redshift dependence of $r_{1/2,\star}/\Rv$ seen in the simulations, motivating the introduction of the additional redshift scaling factor $(1+z)^{-0.2}$.
More recently, analyses of large-box cosmological simulations have shown that, even at fixed mass and redshift, disc sizes follow a similar scaling with concentration \citep{Liang25b,Liang25a}.
This raises the question of whether the concentration dependence in J19 arises  from cosmological assembly effects or instead reflects an imprint of the halo structure itself.

The middle panel of \Fig{predictor_compare} compares our measured stellar half-mass radii to the J19 predictions. 
\footnote{Because the J19 relation depends explicitly on halo concentration and our simulations suite samples three discrete values of concentration, the predicted sizes populate three narrow bands. A small amount of scatter is present because the target values of $c_2$ = 3, 10, and 20 are those measured after the initial relaxation stage. These values were achieved through trial and error and cannot be matched exactly to the intended targets.}
On average, our idealized isolated simulations reproduce the same qualitative concentration trend reported by J19, suggesting that the dependence of galaxy size on concentration does not require a cosmological environment, but instead emerges from the internal dynamics and galaxy-formation physics within virialized haloes.

The right panel of \Fig{predictor_compare} compares the measured 
half-mass radii to a simple predictor based on halo spin \citep{Somerville18},
\be
r_{1/2,\star}\simeq 0.5 \lambda \Rv.
\ee
Both simple empirical models successfully capture the overall trend, but at fixed $c_2$ or $\lambda$, the simulated sizes still span a wide range -- colour-coding by the other quantity reveals  dependence on the other parameter, consistent with what has been shown in \Fig{rsm_interaction}.
We therefore conclude that an accurate prescription for galaxy size requires accounting for multiple halo properties rather than relying on any single parameter.

\subsection{The impact of the galaxy definition}\label{sec:definition}

A practical but often under-emphasized issue in size modeling is that size definitions are not unique. 
Analytic disc models such as Mo98 are formulated for baryonic discs (stars plus cold gas), whereas many simulation and observational comparisons focus on stellar-only size measurements. 
Cold or neutral gas discs are usually thinner and more extended than their stellar counterparts: nearby H\,\textsc{i} surveys find that the ratio of H\,\textsc{i} to optical size is typically $\gtrsim 1$--$2$ \citep[e.g.][]{Broeils97}. 
Simulations likewise produce neutral-gas discs that extended beyond the stellar body \citep[e.g.][]{Marinacci17}. 

\begin{figure}
    \includegraphics[width=\columnwidth]{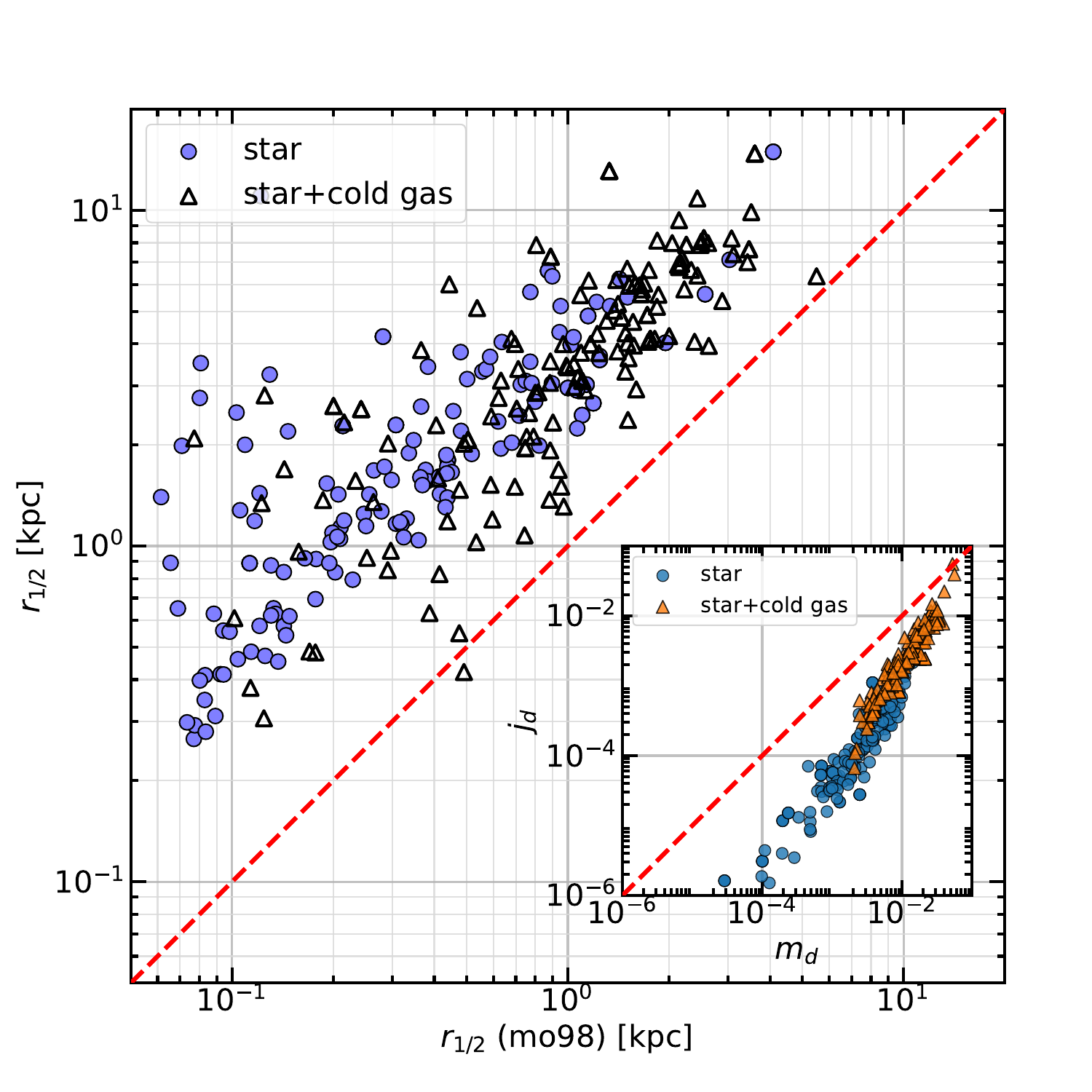}
    \caption{
    Effect of galaxy definition on the size measurement and on the comparison with the \citet{Mo98} (Mo98) predictor. 
    The vertical axis shows the galaxy size measured in the simulations, while the horizontal axis shows the size predicted by the Mo98 model. 
    Symbols indicate measurements using either stars only (blue circles) or total cold baryons (stars plus cold gas with $T<10^{4}$ K; open triangles). 
    The Mo98 predictions are computed with corresponding definitions of the disc mass fraction, $m_\rmd$, and the angular-momentum fraction, $j_\rmd$. 
    The red dashed line marks equality. 
    \quad
    \emph{Inset}: angular-momentum fraction $j_\rmd$ versus mass fraction $m_{\rm d}$ for the two galaxy definitions, with the dashed red line indicating $j_\rmd=m_\rmd$. 
    \quad 
    Including cold gas shifts systems closer to $j_{\rm d}=m_{\rm d}$, modestly improving the agreement between the Mo98 predictions and the measured sizes.
    }
    \label{fig:cold_definition}
\end{figure}

\Fig{cold_definition} quantifies the impact of defining the galaxy using stars alone versus stars plus cold gas. 
The half-mass radius of the cold baryonic component is systematically larger than the stellar-only radius, although the offset is modest compared to the full scatter of the sample. 
This definition-dependent difference affects the performance of the Mo98 model, in which the predicted disc scale depends explicitly on the angular-momentum retention factor, $f_j = j_{\rm d}/m_{\rm d}$.

The inset of \Fig{cold_definition} shows that, under a stellar-only definition, most systems lie well below the $j_{\rm d} = m_{\rm d}$ relation. 
When cold gas is included, however, their locations shift toward $j_{\rm d} = m_{\rm d}$, indicating more efficient angular-momentum retention. 
This behavior is consistent with the expectation that a significant fraction of a disc galaxy’s angular momentum resides in the cold gaseous component \citep[e.g.][]{Sales10}. 
As a result, the radii predicted by the Mo98 model increase, and the agreement with the measured sizes improves modestly when the galaxy is defined to include both stars and cold gas.

\subsection{Limitations of the isolated setup}\label{sec:limitations}

A central limitation of our simulation suite is that it does not follow the hierarchical assembly of galaxies in a cosmological environment.
In a $\Lambda$CDM context, the mass, concentration, shape, and angular momentum of a halo all evolve with time, while the baryons that eventually form the galaxy are supplied through a combination of smooth accretion and mergers.
The angular-momentum direction of the incoming baryons can therefore vary from one accretion episode to another.
Cosmological simulations have shown that this time dependence is important: discs tend to form when the angular momentum of newly accreted gas is coherently aligned with that of the existing galaxy, whereas misaligned accretion can build kinematically distinct stellar components and reduce the net rotational support \citep{Sales12}.
\citet{Zavala16} further showed that the angular-momentum evolution of stars and cold gas is linked to different parts of the halo assembly history, with old stars more closely tracing the inner dark-matter halo and younger stars being more connected to recently accreted high-angular-momentum gas.

Our isolated experiments deliberately remove these complications.
Each run begins from a relaxed halo and gaseous atmosphere with a prescribed structure and a single imposed angular-momentum profile, and the subsequent evolution occurs without mergers or filamentary inflow.
The results should therefore be interpreted as a controlled response experiment: they quantify how galaxy size responds to halo structure and baryon content at fixed initial conditions, rather than predicting the full population scatter produced by cosmological assembly histories. Due to these limitations, we caution against directly comparing our results with studies of galaxy morphology in cosmological simulations. 

It is worth noting that \citet{GarrisonKimmel18} found, for Milky Way-mass \texttt{FIRE-2} galaxies, that the present-day stellar morphology is only weakly correlated with the present-day halo spin (or other instantaneous halo properties), but is more strongly correlated with the spin of the halo gas at the epoch when the galaxy had assembled half of its final stellar mass.
In a sense, this result is consistent with our finding of a significant spin correlation here, as our variation of halo spin may be viewed as mimicking changes in the angular-momentum reservoir at formation time.

\section{Conclusions}\label{sec:conclusion}

In this paper, we have used a suite of controlled, idealized galaxy simulations to isolate how dark-matter halo structural parameters and the initial baryonic mass fraction regulate galaxy size at fixed halo mass, $\Mv=10^{11}\Msun$. 
We have run and analyzed 132 \GIZMO+\FIRE realizations spanning a wide range in halo concentration $c_2$, inner density slope $s_1$, spin $\lambda$, and baryon fraction $f_{\rm b}$. 
For each system, we measured the 3D stellar half-mass radius $r_{1/2,\star}$ and quantified its dependence on halo properties using complementary approaches, including a quadratic response-surface model and a non-parametric random-forest feature-importance ranking. 

Across all diagnostics, halo concentration $c_2$ emerges as the most informative predictor of galaxy size $r_{1/2,\star}$. 
Halo spin $\lambda$ provides the next-largest contribution, while the inner-slope $s_1$ exerts a weaker but still significant influence. 
The baryon fraction $f_{\rm b}$ plays a distinct role: rather than acting as a simple, monotonic regulator of size, it primarily modulates how sensitively galaxy size responds to halo structural parameters.
In particular, baryon fractions approaching the cosmic mean lead to centrally concentrated star formation and compact galaxies, despite the strong stellar feedback implemented in  \FIRE, whereas lower $\fb$ values favor extended, disc-like systems.

We further revisited widely used analytic and empirical size predictors by comparing their predictions to our controlled simulations. 
The \citet{Mo98} model reproduces the overall scaling of galaxy size remarkably well but systematically underpredicts  sizes. 
This is primarily because the simulated galaxies exhibit low angular-momentum retention factors, $f_j\equiv j_{\rm d}/m_{\rm d} \ll 1$.
This suggests that galaxy formation preferentially draws on low-angular-momentum baryons and that stellar feedback redistributes angular momentum. 

The empirical, concentration-based predictor of \citet{Jiang19} successfully captures the qualitative trend that galaxies in lower-concentration haloes are larger on average, $r_{1/2,\star}\simeq 0.02 (c_2/10)^{-0.7} \Rv$. 
Our idealized simulations demonstrate that this dependence does not require a cosmological environment, but instead arises naturally from the internal dynamics and galaxy-formation physics within virialized haloes.  
However, the large scatter in $r_{1/2,\star}$ at fixed $c_2$ underscores the limitations of single-parameter size prescriptions and highlights the need to account for additional halo properties such as spin, inner structure, and baryon content.
Similarly, a simple halo-spin-based predictor of $r_{1/2,\star}\simeq 0.5 \lambda \Rv$ \citep{Somerville18} also successfully captures the overall trend that galaxies in higher-spin haloes are larger, but leaves a large scatter at fixed spin that requires additional modeling.

Finally, we showed that galaxy definition itself matters. 
Defining the galaxy as stars plus cold gas ($T<10^4\,{\rm K}$) will systematically increase the half-mass radius relative to a stellar-only definition, consistent with the expectation that cold gas discs are more extended. 
Including cold gas moves systems closer to $j_{\rm d}=m_{\rm d}$, improving angular-momentum retention for the baryonic component and the agreement between Mo98 and the measured sizes.

This work establishes a controlled baseline for understanding how secondary halo properties regulate galaxy sizes at the massive-dwarf scale. 
Future papers in this series will extend this framework to additional halo masses and to other aspects of galaxy morphology, including substructure driven by feedback, such as clumps, cavities, and asymmetries, and will quantify how these features depend on halo structural conditions.

\section*{Acknowledgements}
We appreciate helpful discussions with Miki Yohei, Hui Li, Florent Renaud, Xi Kang, Suoqing Ji, Houzun Chen, and Jinning Liang.
We thank Philip F. Hopkins for sharing the \GIZMO simulation code with the \FIRE prescriptions. 
FJ acknowledges support by the National Natural Science Foundation of China (NSFC, Grant No. 12473007) and China Manned Space Program with Grant No. CMS-CSST-2025-A03.
We acknowledge the High-performance Computing Platform of Peking University for providing computational resources and support.

\section*{Data Availability}
The data products and the analysis pipeline are available upon request. 
\bibliographystyle{mnras}
\bibliography{mnras} 

@ARTICLE{Peebles69,
       author = {{Peebles}, P.~J.~E.},
        title = "{Origin of the Angular Momentum of Galaxies}",
      journal = {\apj},
         year = 1969,
        month = feb,
       volume = {155},
        pages = {393},
          doi = {10.1086/149876},
       adsurl = {https://ui.adsabs.harvard.edu/abs/1969ApJ...155..393P}
}

@ARTICLE{Press74,
       author = {{Press}, William H. and {Schechter}, Paul},
        title = "{Formation of Galaxies and Clusters of Galaxies by Self-Similar Gravitational Condensation}",
      journal = {\apj},
         year = 1974,
        month = feb,
       volume = {187},
        pages = {425-438},
          doi = {10.1086/152650},
       adsurl = {https://ui.adsabs.harvard.edu/abs/1974ApJ...187..425P}
}

@ARTICLE{White78,
       author = {{White}, S.~D.~M. and {Rees}, M.~J.},
        title = "{Core condensation in heavy halos: a two-stage theory for galaxy formation and clustering.}",
      journal = {\mnras},
         year = 1978,
        month = may,
       volume = {183},
        pages = {341-358},
          doi = {10.1093/mnras/183.3.341},
       adsurl = {https://ui.adsabs.harvard.edu/abs/1978MNRAS.183..341W}
}

@ARTICLE{Fall80,
       author = {{Fall}, S.~M. and {Efstathiou}, G.},
        title = "{Formation and rotation of disc galaxies with haloes.}",
      journal = {\mnras},
         year = 1980,
        month = oct,
       volume = {193},
        pages = {189-206},
          doi = {10.1093/mnras/193.2.189},
       adsurl = {https://ui.adsabs.harvard.edu/abs/1980MNRAS.193..189F}
}

@ARTICLE{Blumenthal84,
       author = {{Blumenthal}, G.~R. and {Faber}, S.~M. and {Primack}, J.~R. and {Rees}, M.~J.},
        title = "{Formation of galaxies and large-scale structure with cold dark matter.}",
      journal = {\nat},
         year = 1984,
        month = oct,
       volume = {311},
        pages = {517-525},
          doi = {10.1038/311517a0},
       adsurl = {https://ui.adsabs.harvard.edu/abs/1984Natur.311..517B}
}

@ARTICLE{LaceyCole93,
       author = {{Lacey}, Cedric and {Cole}, Shaun},
        title = "{Merger rates in hierarchical models of galaxy formation}",
      journal = {\mnras},
         year = 1993,
        month = jun,
       volume = {262},
       number = {3},
        pages = {627-649},
          doi = {10.1093/mnras/262.3.627},
       adsurl = {https://ui.adsabs.harvard.edu/abs/1993MNRAS.262..627L}
}

@ARTICLE{Navarro96,
       author = {{Navarro}, Julio F. and {Frenk}, Carlos S. and {White}, Simon D.~M.},
        title = "{The Structure of Cold Dark Matter Halos}",
      journal = {\apj},
         year = 1996,
        month = may,
       volume = {462},
        pages = {563},
          doi = {10.1086/177173},
archivePrefix = {arXiv},
       eprint = {astro-ph/9508025},
 primaryClass = {astro-ph},
       adsurl = {https://ui.adsabs.harvard.edu/abs/1996ApJ...462..563N}
}

@ARTICLE{Broeils97,
       author = {{Broeils}, A.~H. and {Rhee}, M.-H.},
        title = "{Short 21-cm WSRT observations of spiral and irregular galaxies. HI properties.}",
      journal = {\aap},
         year = 1997,
        month = aug,
       volume = {324},
        pages = {877-887},
       adsurl = {https://ui.adsabs.harvard.edu/abs/1997A&A...324..877B}
}

@ARTICLE{Navarro97,
       author = {{Navarro}, Julio F. and {Frenk}, Carlos S. and {White}, Simon D.~M.},
        title = "{A Universal Density Profile from Hierarchical Clustering}",
      journal = {\apj},
         year = 1997,
        month = dec,
       volume = {490},
       number = {2},
        pages = {493-508},
          doi = {10.1086/304888},
archivePrefix = {arXiv},
       eprint = {astro-ph/9611107},
 primaryClass = {astro-ph},
       adsurl = {https://ui.adsabs.harvard.edu/abs/1997ApJ...490..493N}
}

@ARTICLE{Mo98,
       author = {{Mo}, H.~J. and {Mao}, Shude and {White}, Simon D.~M.},
        title = "{The formation of galactic discs}",
      journal = {\mnras},
         year = 1998,
        month = apr,
       volume = {295},
       number = {2},
        pages = {319-336},
          doi = {10.1046/j.1365-8711.1998.01227.x},
archivePrefix = {arXiv},
       eprint = {astro-ph/9707093},
 primaryClass = {astro-ph},
       adsurl = {https://ui.adsabs.harvard.edu/abs/1998MNRAS.295..319M}
}

@ARTICLE{Ferland98,
       author = {{Ferland}, G.~J. and {Korista}, K.~T. and {Verner}, D.~A. and {Ferguson}, J.~W. and {Kingdon}, J.~B. and {Verner}, E.~M.},
        title = "{CLOUDY 90: Numerical Simulation of Plasmas and Their Spectra}",
      journal = {\pasp},
         year = 1998,
        month = jul,
       volume = {110},
       number = {749},
        pages = {761-778},
          doi = {10.1086/316190},
       adsurl = {https://ui.adsabs.harvard.edu/abs/1998PASP..110..761F}
}

@ARTICLE{Breiman01,
       author = {{Breiman}, Leo},
        title = "{Random Forests.}",
      journal = {Machine Learning},
         year = 2001,
        month = jan,
       volume = {45},
        pages = {5-32},
          doi = {10.1023/A:1010933404324},
       adsurl = {https://ui.adsabs.harvard.edu/abs/2001MachL..45....5B}
}

@ARTICLE{Kroupa01,
       author = {{Kroupa}, Pavel},
        title = "{On the variation of the initial mass function}",
      journal = {\mnras},
         year = 2001,
        month = apr,
       volume = {322},
       number = {2},
        pages = {231-246},
          doi = {10.1046/j.1365-8711.2001.04022.x},
archivePrefix = {arXiv},
       eprint = {astro-ph/0009005},
 primaryClass = {astro-ph},
       adsurl = {https://ui.adsabs.harvard.edu/abs/2001MNRAS.322..231K}
}

@ARTICLE{Bullock01b,
       author = {{Bullock}, J.~S. and {Dekel}, A. and {Kolatt}, T.~S. and {Kravtsov}, A.~V. and {Klypin}, A.~A. and {Porciani}, C. and {Primack}, J.~R.},
        title = "{A Universal Angular Momentum Profile for Galactic Halos}",
      journal = {\apj},
         year = 2001,
        month = jul,
       volume = {555},
       number = {1},
        pages = {240-257},
          doi = {10.1086/321477},
archivePrefix = {arXiv},
       eprint = {astro-ph/0011001},
 primaryClass = {astro-ph},
       adsurl = {https://ui.adsabs.harvard.edu/abs/2001ApJ...555..240B}
}

@ARTICLE{Zhao03,
       author = {{Zhao}, D.~H. and {Mo}, H.~J. and {Jing}, Y.~P. and {B{\"o}rner}, G.},
        title = "{The growth and structure of dark matter haloes}",
      journal = {\mnras},
         year = 2003,
        month = feb,
       volume = {339},
       number = {1},
        pages = {12-24},
          doi = {10.1046/j.1365-8711.2003.06135.x},
archivePrefix = {arXiv},
       eprint = {astro-ph/0204108},
 primaryClass = {astro-ph},
       adsurl = {https://ui.adsabs.harvard.edu/abs/2003MNRAS.339...12Z}
}

@ARTICLE{ShethTormen04,
       author = {{Sheth}, Ravi K. and {Tormen}, Giuseppe},
        title = "{Formation times and masses of dark matter haloes}",
      journal = {\mnras},
         year = 2004,
        month = apr,
       volume = {349},
       number = {4},
        pages = {1464-1468},
          doi = {10.1111/j.1365-2966.2004.07622.x},
archivePrefix = {arXiv},
       eprint = {astro-ph/0402055},
 primaryClass = {astro-ph},
       adsurl = {https://ui.adsabs.harvard.edu/abs/2004MNRAS.349.1464S}
}

@ARTICLE{Springel05b,
       author = {{Springel}, Volker and {White}, Simon D.~M. and {Jenkins}, Adrian and {Frenk}, Carlos S. and {Yoshida}, Naoki and {Gao}, Liang and {Navarro}, Julio and {Thacker}, Robert and {Croton}, Darren and {Helly}, John and {Peacock}, John A. and {Cole}, Shaun and {Thomas}, Peter and {Couchman}, Hugh and {Evrard}, August and {Colberg}, J{\"o}rg and {Pearce}, Frazer},
        title = "{Simulations of the formation, evolution and clustering of galaxies and quasars}",
      journal = {\nat},
         year = 2005,
        month = jun,
       volume = {435},
       number = {7042},
        pages = {629-636},
          doi = {10.1038/nature03597},
archivePrefix = {arXiv},
       eprint = {astro-ph/0504097},
 primaryClass = {astro-ph},
       adsurl = {https://ui.adsabs.harvard.edu/abs/2005Natur.435..629S}
}

@ARTICLE{Springel05a,
       author = {{Springel}, Volker},
        title = "{The cosmological simulation code GADGET-2}",
      journal = {\mnras},
         year = 2005,
        month = dec,
       volume = {364},
       number = {4},
        pages = {1105-1134},
          doi = {10.1111/j.1365-2966.2005.09655.x},
archivePrefix = {arXiv},
       eprint = {astro-ph/0505010},
 primaryClass = {astro-ph},
       adsurl = {https://ui.adsabs.harvard.edu/abs/2005MNRAS.364.1105S}
}

@ARTICLE{Li08,
       author = {{Li}, Yun and {Mo}, H.~J. and {Gao}, L.},
        title = "{On halo formation times and assembly bias}",
      journal = {\mnras},
         year = 2008,
        month = sep,
       volume = {389},
       number = {3},
        pages = {1419-1426},
          doi = {10.1111/j.1365-2966.2008.13667.x},
archivePrefix = {arXiv},
       eprint = {0803.2250},
 primaryClass = {astro-ph},
       adsurl = {https://ui.adsabs.harvard.edu/abs/2008MNRAS.389.1419L}
}

@ARTICLE{Wiersma09,
       author = {{Wiersma}, Robert P.~C. and {Schaye}, Joop and {Theuns}, Tom and {Dalla Vecchia}, Claudio and {Tornatore}, Luca},
        title = "{Chemical enrichment in cosmological, smoothed particle hydrodynamics simulations}",
      journal = {\mnras},
         year = 2009,
        month = oct,
       volume = {399},
       number = {2},
        pages = {574-600},
          doi = {10.1111/j.1365-2966.2009.15331.x},
archivePrefix = {arXiv},
       eprint = {0902.1535},
 primaryClass = {astro-ph.CO},
       adsurl = {https://ui.adsabs.harvard.edu/abs/2009MNRAS.399..574W}
}

@ARTICLE{Sales10,
       author = {{Sales}, Laura V. and {Navarro}, Julio F. and {Schaye}, Joop and {Dalla Vecchia}, Claudio and {Springel}, Volker and {Booth}, C.~M.},
        title = "{Feedback and the structure of simulated galaxies at redshift z= 2}",
      journal = {\mnras},
         year = 2010,
        month = dec,
       volume = {409},
       number = {4},
        pages = {1541-1556},
          doi = {10.1111/j.1365-2966.2010.17391.x},
archivePrefix = {arXiv},
       eprint = {1004.5386},
 primaryClass = {astro-ph.CO},
       adsurl = {https://ui.adsabs.harvard.edu/abs/2010MNRAS.409.1541S}
}

@article{scikit-learn,
  title   = {Scikit-learn: Machine Learning in {P}ython},
  author  = {Pedregosa, F. and Varoquaux, G. and Gramfort, A. and Michel, V.
             and Thirion, B. and Grisel, O. and Blondel, M. and Prettenhofer, P.
             and Weiss, R. and Dubourg, V. and Vanderplas, J. and Passos, A. and
             Cournapeau, D. and Brucher, M. and Perrot, M. and Duchesnay, E.},
  journal = {Journal of Machine Learning Research},
  volume  = {12},
  pages   = {2825--2830},
  year    = {2011}
}

@ARTICLE{Turk11,
       author = {{Turk}, Matthew J. and {Smith}, Britton D. and {Oishi}, Jeffrey S. and {Skory}, Stephen and {Skillman}, Samuel W. and {Abel}, Tom and {Norman}, Michael L.},
        title = "{yt: A Multi-code Analysis Toolkit for Astrophysical Simulation Data}",
      journal = {\apjs},
         year = 2011,
        month = jan,
       volume = {192},
       number = {1},
          eid = {9},
        pages = {9},
          doi = {10.1088/0067-0049/192/1/9},
archivePrefix = {arXiv},
       eprint = {1011.3514},
 primaryClass = {astro-ph.IM},
       adsurl = {https://ui.adsabs.harvard.edu/abs/2011ApJS..192....9T}
}

@ARTICLE{Sales12,
       author = {{Sales}, Laura V. and {Navarro}, Julio F. and {Theuns}, Tom and {Schaye}, Joop and {White}, Simon D.~M. and {Frenk}, Carlos S. and {Crain}, Robert A. and {Dalla Vecchia}, Claudio},
        title = "{The origin of discs and spheroids in simulated galaxies}",
      journal = {\mnras},
         year = 2012,
        month = jun,
       volume = {423},
       number = {2},
        pages = {1544-1555},
          doi = {10.1111/j.1365-2966.2012.20975.x},
archivePrefix = {arXiv},
       eprint = {1112.2220},
 primaryClass = {astro-ph.CO},
       adsurl = {https://ui.adsabs.harvard.edu/abs/2012MNRAS.423.1544S}
}

@ARTICLE{Zavala16,
       author = {{Zavala}, Jes{\'u}s and {Frenk}, Carlos S. and {Bower}, Richard and {Schaye}, Joop and {Theuns}, Tom and {Crain}, Robert A. and {Trayford}, James W. and {Schaller}, Matthieu and {Furlong}, Michelle},
        title = "{The link between the assembly of the inner dark matter halo and the angular momentum evolution of galaxies in the EAGLE simulation}",
      journal = {\mnras},
         year = 2016,
        month = aug,
       volume = {460},
       number = {4},
        pages = {4466-4482},
          doi = {10.1093/mnras/stw1286},
       adsurl = {https://ui.adsabs.harvard.edu/abs/2016MNRAS.460.4466Z}
}

@ARTICLE{Moster13,
       author = {{Moster}, Benjamin P. and {Naab}, Thorsten and {White}, Simon D.~M.},
        title = "{Galactic star formation and accretion histories from matching galaxies to dark matter haloes}",
      journal = {\mnras},
         year = 2013,
        month = feb,
       volume = {428},
       number = {4},
        pages = {3121-3138},
          doi = {10.1093/mnras/sts261},
archivePrefix = {arXiv},
       eprint = {1205.5807},
 primaryClass = {astro-ph.CO},
       adsurl = {https://ui.adsabs.harvard.edu/abs/2013MNRAS.428.3121M}
}

@ARTICLE{Kravtsov13,
       author = {{Kravtsov}, Andrey V.},
        title = "{The Size-Virial Radius Relation of Galaxies}",
      journal = {\apjl},
         year = 2013,
        month = feb,
       volume = {764},
       number = {2},
          eid = {L31},
        pages = {L31},
          doi = {10.1088/2041-8205/764/2/L31},
archivePrefix = {arXiv},
       eprint = {1212.2980},
 primaryClass = {astro-ph.CO},
       adsurl = {https://ui.adsabs.harvard.edu/abs/2013ApJ...764L..31K}
}

@misc{Pontzen13,
       author = {{Pontzen}, Andrew and {Ro{\v{s}}kar}, Rok and {Stinson}, Greg and {Woods}, Rory},
        title = "{pynbody: N-Body/SPH analysis for python}",
 howpublished = {Astrophysics Source Code Library, record ascl:1305.002},
         year = 2013,
        month = may,
          eid = {ascl:1305.002},
archivePrefix = {ascl},
       eprint = {1305.002},
       adsurl = {https://ui.adsabs.harvard.edu/abs/2013ascl.soft05002P}
}

@ARTICLE{Behroozi13,
       author = {{Behroozi}, Peter S. and {Wechsler}, Risa H. and {Conroy}, Charlie},
        title = "{The Average Star Formation Histories of Galaxies in Dark Matter Halos from z = 0-8}",
      journal = {\apj},
         year = 2013,
        month = jun,
       volume = {770},
       number = {1},
          eid = {57},
        pages = {57},
          doi = {10.1088/0004-637X/770/1/57},
archivePrefix = {arXiv},
       eprint = {1207.6105},
 primaryClass = {astro-ph.CO},
       adsurl = {https://ui.adsabs.harvard.edu/abs/2013ApJ...770...57B}
}

@ARTICLE{KimLee13,
       author = {{Kim}, Ji-hoon and {Lee}, Jounghun},
        title = "{How does the surface density and size of disc galaxies measured in hydrodynamic simulations correlate with the halo spin parameter?}",
      journal = {\mnras},
         year = 2013,
        month = jun,
       volume = {432},
       number = {2},
        pages = {1701-1708},
          doi = {10.1093/mnras/stt632},
archivePrefix = {arXiv},
       eprint = {1210.8321},
 primaryClass = {astro-ph.GA},
       adsurl = {https://ui.adsabs.harvard.edu/abs/2013MNRAS.432.1701K}
}

@ARTICLE{Hopkins13,
       author = {{Hopkins}, Philip F. and {Narayanan}, Desika and {Murray}, Norman},
        title = "{The meaning and consequences of star formation criteria in galaxy models with resolved stellar feedback}",
      journal = {\mnras},
         year = 2013,
        month = jul,
       volume = {432},
       number = {4},
        pages = {2647-2653},
          doi = {10.1093/mnras/stt723},
archivePrefix = {arXiv},
       eprint = {1303.0285},
 primaryClass = {astro-ph.CO},
       adsurl = {https://ui.adsabs.harvard.edu/abs/2013MNRAS.432.2647H}
}

@ARTICLE{Cintio14,
       author = {{Di Cintio}, Arianna and {Brook}, Chris B. and {Macci{\`o}}, Andrea V. and {Stinson}, Greg S. and {Knebe}, Alexander and {Dutton}, Aaron A. and {Wadsley}, James},
        title = "{The dependence of dark matter profiles on the stellar-to-halo mass ratio: a prediction for cusps versus cores}",
      journal = {\mnras},
         year = 2014,
        month = jan,
       volume = {437},
       number = {1},
        pages = {415-423},
          doi = {10.1093/mnras/stt1891},
archivePrefix = {arXiv},
       eprint = {1306.0898},
 primaryClass = {astro-ph.CO},
       adsurl = {https://ui.adsabs.harvard.edu/abs/2014MNRAS.437..415D}
}

@ARTICLE{DuttonMaccio14,
       author = {{Dutton}, Aaron A. and {Macci{\`o}}, Andrea V.},
        title = "{Cold dark matter haloes in the Planck era: evolution of structural parameters for Einasto and NFW profiles}",
      journal = {\mnras},
         year = 2014,
        month = jul,
       volume = {441},
       number = {4},
        pages = {3359-3374},
          doi = {10.1093/mnras/stu742},
archivePrefix = {arXiv},
       eprint = {1402.7073},
 primaryClass = {astro-ph.CO},
       adsurl = {https://ui.adsabs.harvard.edu/abs/2014MNRAS.441.3359D}
}

@ARTICLE{DiemerKravtsov15,
       author = {{Diemer}, Benedikt and {Kravtsov}, Andrey V.},
        title = "{A Universal Model for Halo Concentrations}",
      journal = {\apj},
         year = 2015,
        month = jan,
       volume = {799},
       number = {1},
          eid = {108},
        pages = {108},
          doi = {10.1088/0004-637X/799/1/108},
archivePrefix = {arXiv},
       eprint = {1407.4730},
 primaryClass = {astro-ph.CO},
       adsurl = {https://ui.adsabs.harvard.edu/abs/2015ApJ...799..108D}
}

@ARTICLE{Hopkins15,
       author = {{Hopkins}, Philip F.},
        title = "{A new class of accurate, mesh-free hydrodynamic simulation methods}",
      journal = {\mnras},
         year = 2015,
        month = jun,
       volume = {450},
       number = {1},
        pages = {53-110},
          doi = {10.1093/mnras/stv195},
archivePrefix = {arXiv},
       eprint = {1409.7395},
 primaryClass = {astro-ph.CO},
       adsurl = {https://ui.adsabs.harvard.edu/abs/2015MNRAS.450...53H}
}

@ARTICLE{Oman15,
       author = {{Oman}, Kyle A. and {Navarro}, Julio F. and {Fattahi}, Azadeh and {Frenk}, Carlos S. and {Sawala}, Till and {White}, Simon D.~M. and {Bower}, Richard and {Crain}, Robert A. and {Furlong}, Michelle and {Schaller}, Matthieu and {Schaye}, Joop and {Theuns}, Tom},
        title = "{The unexpected diversity of dwarf galaxy rotation curves}",
      journal = {\mnras},
         year = 2015,
        month = oct,
       volume = {452},
       number = {4},
        pages = {3650-3665},
          doi = {10.1093/mnras/stv1504},
archivePrefix = {arXiv},
       eprint = {1504.01437},
 primaryClass = {astro-ph.GA},
       adsurl = {https://ui.adsabs.harvard.edu/abs/2015MNRAS.452.3650O}
}

@ARTICLE{Wang15NIHAO,
       author = {{Wang}, Liang and {Dutton}, Aaron A. and {Stinson}, Gregory S. and {Macci{\`o}}, Andrea V. and {Penzo}, Camilla and {Kang}, Xi and {Keller}, Ben W. and {Wadsley}, James},
        title = "{NIHAO project - I. Reproducing the inefficiency of galaxy formation across cosmic time with a large sample of cosmological hydrodynamical simulations}",
      journal = {\mnras},
         year = 2015,
        month = nov,
       volume = {454},
       number = {1},
        pages = {83-94},
          doi = {10.1093/mnras/stv1937},
archivePrefix = {arXiv},
       eprint = {1503.04818},
 primaryClass = {astro-ph.GA},
       adsurl = {https://ui.adsabs.harvard.edu/abs/2015MNRAS.454...83W}
}

@book{Myers2016RSM,
  title     = {Response Surface Methodology: Process and Product Optimization Using Designed Experiments},
  author    = {Myers, Raymond H. and Montgomery, Douglas C. and Anderson-Cook, Christine M.},
  year      = {2016},
  month     = feb,
  edition   = {4},
  publisher = {John Wiley \& Sons},
  series    = {Wiley Series in Probability and Statistics},
  isbn      = {9781118916018},
  url       = {https://books.google.com/books?id=YFSzCgAAQBAJ}
}

@ARTICLE{Marinacci17,
       author = {{Marinacci}, Federico and {Grand}, Robert J.~J. and {Pakmor}, R{\"u}diger and {Springel}, Volker and {G{\'o}mez}, Facundo A. and {Frenk}, Carlos S. and {White}, Simon D.~M.},
        title = "{Properties of H I discs in the Auriga cosmological simulations}",
      journal = {\mnras},
         year = 2017,
        month = apr,
       volume = {466},
       number = {4},
        pages = {3859-3875},
          doi = {10.1093/mnras/stw3366},
archivePrefix = {arXiv},
       eprint = {1610.01594},
 primaryClass = {astro-ph.GA},
       adsurl = {https://ui.adsabs.harvard.edu/abs/2017MNRAS.466.3859M}
}

@ARTICLE{Grand17,
       author = {{Grand}, Robert J.~J. and {G{\'o}mez}, Facundo A. and {Marinacci}, Federico and {Pakmor}, R{\"u}diger and {Springel}, Volker and {Campbell}, David J.~R. and {Frenk}, Carlos S. and {Jenkins}, Adrian and {White}, Simon D.~M.},
        title = "{The Auriga Project: the properties and formation mechanisms of disc galaxies across cosmic time}",
      journal = {\mnras},
         year = 2017,
        month = may,
       volume = {467},
       number = {1},
        pages = {179-207},
          doi = {10.1093/mnras/stx071},
archivePrefix = {arXiv},
       eprint = {1610.01159},
 primaryClass = {astro-ph.GA},
       adsurl = {https://ui.adsabs.harvard.edu/abs/2017MNRAS.467..179G}
}

@ARTICLE{RodriguezGomez17,
       author = {{Rodriguez-Gomez}, Vicente and {Sales}, Laura V. and {Genel}, Shy and {Pillepich}, Annalisa and {Zjupa}, Jolanta and {Nelson}, Dylan and {Griffen}, Brendan and {Torrey}, Paul and {Snyder}, Gregory F. and {Vogelsberger}, Mark and {Springel}, Volker and {Ma}, Chung-Pei and {Hernquist}, Lars},
        title = "{The role of mergers and halo spin in shaping galaxy morphology}",
      journal = {\mnras},
         year = 2017,
        month = may,
       volume = {467},
       number = {3},
        pages = {3083-3098},
          doi = {10.1093/mnras/stx305},
archivePrefix = {arXiv},
       eprint = {1609.09498},
 primaryClass = {astro-ph.GA},
       adsurl = {https://ui.adsabs.harvard.edu/abs/2017MNRAS.467.3083R}
}

@ARTICLE{GarrisonKimmel18,
       author = {{Garrison-Kimmel}, Shea and {Hopkins}, Philip F. and {Wetzel}, Andrew and {El-Badry}, Kareem and {Sanderson}, Robyn E. and {Bullock}, James S. and {Ma}, Xiangcheng and {van de Voort}, Freeke and {Hafen}, Zachary and {Faucher-Gigu{\`e}re}, Claude-Andr{\'e} and {Hayward}, Christopher C. and {Quataert}, Eliot and {Kere{\v{s}}}, Du{\v{s}}an and {Boylan-Kolchin}, Michael},
        title = "{The origin of the diverse morphologies and kinematics of Milky Way-mass galaxies in the FIRE-2 simulations}",
      journal = {\mnras},
         year = 2018,
        month = dec,
       volume = {481},
       number = {3},
        pages = {4133-4157},
          doi = {10.1093/mnras/sty2513},
       adsurl = {https://ui.adsabs.harvard.edu/abs/2018MNRAS.481.4133G}
}

@ARTICLE{Lundberg17,
       author = {{Lundberg}, Scott and {Lee}, Su-In},
        title = "{A Unified Approach to Interpreting Model Predictions}",
      journal = {arXiv e-prints},
         year = 2017,
        month = may,
          eid = {arXiv:1705.07874},
        pages = {arXiv:1705.07874},
          doi = {10.48550/arXiv.1705.07874},
archivePrefix = {arXiv},
       eprint = {1705.07874},
 primaryClass = {cs.AI},
       adsurl = {https://ui.adsabs.harvard.edu/abs/2017arXiv170507874L}
}

@ARTICLE{Somerville18,
       author = {{Somerville}, Rachel S. and {Behroozi}, Peter and {Pandya}, Viraj and {Dekel}, Avishai and {Faber}, S.~M. and {Fontana}, Adriano and {Koekemoer}, Anton M. and {Koo}, David C. and {P{\'e}rez-Gonz{\'a}lez}, P.~G. and {Primack}, Joel R. and {Santini}, Paola and {Taylor}, Edward N. and {van der Wel}, Arjen},
        title = "{The relationship between galaxy and dark matter halo size from z {\ensuremath{\sim}} 3 to the present}",
      journal = {\mnras},
         year = 2018,
        month = jan,
       volume = {473},
       number = {2},
        pages = {2714-2736},
          doi = {10.1093/mnras/stx2040},
archivePrefix = {arXiv},
       eprint = {1701.03526},
 primaryClass = {astro-ph.GA},
       adsurl = {https://ui.adsabs.harvard.edu/abs/2018MNRAS.473.2714S}
}

@ARTICLE{Clauwens18,
       author = {{Clauwens}, Bart and {Schaye}, Joop and {Franx}, Marijn and {Bower}, Richard G.},
        title = "{The three phases of galaxy formation}",
      journal = {\mnras},
         year = 2018,
        month = aug,
       volume = {478},
       number = {3},
        pages = {3994-4009},
          doi = {10.1093/mnras/sty1229}
}

@ARTICLE{Wechsler18,
       author = {{Wechsler}, Risa H. and {Tinker}, Jeremy L.},
        title = "{The Connection Between Galaxies and Their Dark Matter Halos}",
      journal = {\araa},
         year = 2018,
        month = sep,
       volume = {56},
        pages = {435-487},
          doi = {10.1146/annurev-astro-081817-051756},
archivePrefix = {arXiv},
       eprint = {1804.03097},
 primaryClass = {astro-ph.GA},
       adsurl = {https://ui.adsabs.harvard.edu/abs/2018ARA&A..56..435W}
}

@ARTICLE{Hopkins18,
       author = {{Hopkins}, Philip F. and {Wetzel}, Andrew and {Kere{\v{s}}}, Du{\v{s}}an and {Faucher-Gigu{\`e}re}, Claude-Andr{\'e} and {Quataert}, Eliot and {Boylan-Kolchin}, Michael and {Murray}, Norman and {Hayward}, Christopher C. and {Garrison-Kimmel}, Shea and {Hummels}, Cameron and {Feldmann}, Robert and {Torrey}, Paul and {Ma}, Xiangcheng and {Angl{\'e}s-Alc{\'a}zar}, Daniel and {Su}, Kung-Yi and {Orr}, Matthew and {Schmitz}, Denise and {Escala}, Ivanna and {Sanderson}, Robyn and {Grudi{\'c}}, Michael Y. and {Hafen}, Zachary and {Kim}, Ji-Hoon and {Fitts}, Alex and {Bullock}, James S. and {Wheeler}, Coral and {Chan}, T.~K. and {Elbert}, Oliver D. and {Narayanan}, Desika},
        title = "{FIRE-2 simulations: physics versus numerics in galaxy formation}",
      journal = {\mnras},
         year = 2018,
        month = oct,
       volume = {480},
       number = {1},
        pages = {800-863},
          doi = {10.1093/mnras/sty1690},
archivePrefix = {arXiv},
       eprint = {1702.06148},
 primaryClass = {astro-ph.GA},
       adsurl = {https://ui.adsabs.harvard.edu/abs/2018MNRAS.480..800H}
}

@ARTICLE{Fielding18,
       author = {{Fielding}, Drummond and {Quataert}, Eliot and {Martizzi}, Davide},
        title = "{Clustered supernovae drive powerful galactic winds after superbubble breakout}",
      journal = {\mnras},
         year = 2018,
        month = dec,
       volume = {481},
       number = {3},
        pages = {3325-3347},
          doi = {10.1093/mnras/sty2466},
archivePrefix = {arXiv},
       eprint = {1807.08758},
 primaryClass = {astro-ph.GA},
       adsurl = {https://ui.adsabs.harvard.edu/abs/2018MNRAS.481.3325F}
}

@ARTICLE{Tacchella19,
       author = {{Tacchella}, Sandro and {Diemer}, Benedikt and {Hernquist}, Lars and {Genel}, Shy and {Marinacci}, Federico and {Nelson}, Dylan and {Pillepich}, Annalisa and {Rodriguez-Gomez}, Vicente and {Sales}, Laura V. and {Springel}, Volker and {Vogelsberger}, Mark},
        title = "{Morphology and star formation in IllustrisTNG: the build-up of spheroids and discs}",
      journal = {\mnras},
         year = 2019,
        month = aug,
       volume = {487},
       number = {4},
        pages = {5416-5440},
          doi = {10.1093/mnras/stz1657}
}

@ARTICLE{Bose19,
       author = {{Bose}, Sownak and {Frenk}, Carlos S. and {Jenkins}, Adrian and {Fattahi}, Azadeh and {G{\'o}mez}, Facundo A. and {Grand}, Robert J.~J. and {Marinacci}, Federico and {Navarro}, Julio F. and {Oman}, Kyle A. and {Pakmor}, R{\"u}diger and {Schaye}, Joop and {Simpson}, Christine M. and {Springel}, Volker},
        title = "{No cores in dark matter-dominated dwarf galaxies with bursty star formation histories}",
      journal = {\mnras},
         year = 2019,
        month = jul,
       volume = {486},
       number = {4},
        pages = {4790-4804},
          doi = {10.1093/mnras/stz1168},
archivePrefix = {arXiv},
       eprint = {1810.03635},
 primaryClass = {astro-ph.GA},
       adsurl = {https://ui.adsabs.harvard.edu/abs/2019MNRAS.486.4790B}
}

@ARTICLE{Jiang19,
       author = {{Jiang}, Fangzhou and {Dekel}, Avishai and {Kneller}, Omer and {Lapiner}, Sharon and {Ceverino}, Daniel and {Primack}, Joel R. and {Faber}, Sandra M. and {Macci{\`o}}, Andrea V. and {Dutton}, Aaron A. and {Genel}, Shy and {Somerville}, Rachel S.},
        title = "{Is the dark-matter halo spin a predictor of galaxy spin and size?}",
      journal = {\mnras},
         year = 2019,
        month = oct,
       volume = {488},
       number = {4},
        pages = {4801-4815},
          doi = {10.1093/mnras/stz1952},
archivePrefix = {arXiv},
       eprint = {1804.07306},
 primaryClass = {astro-ph.GA},
       adsurl = {https://ui.adsabs.harvard.edu/abs/2019MNRAS.488.4801J}
}

@ARTICLE{Hafen19,
       author = {{Hafen}, Zachary and {Faucher-Gigu{\`e}re}, Claude-Andr{\'e} and {Angl{\'e}s-Alc{\'a}zar}, Daniel and {Stern}, Jonathan and {Kere{\v{s}}}, Du{\v{s}}an and {Hummels}, Cameron and {Esmerian}, Clarke and {Garrison-Kimmel}, Shea and {El-Badry}, Kareem and {Wetzel}, Andrew and {Chan}, T.~K. and {Hopkins}, Philip F. and {Murray}, Norman},
        title = "{The origins of the circumgalactic medium in the FIRE simulations}",
      journal = {\mnras},
         year = 2019,
        month = sep,
       volume = {488},
       number = {1},
        pages = {1248-1272},
          doi = {10.1093/mnras/stz1773},
archivePrefix = {arXiv},
       eprint = {1811.11753},
 primaryClass = {astro-ph.GA},
       adsurl = {https://ui.adsabs.harvard.edu/abs/2019MNRAS.488.1248H}
}

@ARTICLE{Hopkins20,
       author = {{Hopkins}, Philip F. and {Grudi{\'c}}, Michael Y. and {Wetzel}, Andrew and {Kere{\v{s}}}, Du{\v{s}}an and {Faucher-Gigu{\`e}re}, Claude-Andr{\'e} and {Ma}, Xiangcheng and {Murray}, Norman and {Butcher}, Nathan},
        title = "{Radiative stellar feedback in galaxy formation: Methods and physics}",
      journal = {\mnras},
         year = 2020,
        month = jan,
       volume = {491},
       number = {3},
        pages = {3702-3729},
          doi = {10.1093/mnras/stz3129},
archivePrefix = {arXiv},
       eprint = {1811.12462},
 primaryClass = {astro-ph.GA},
       adsurl = {https://ui.adsabs.harvard.edu/abs/2020MNRAS.491.3702H}
}

@misc{Wetzel20,
       author = {{Wetzel}, Andrew and {Garrison-Kimmel}, Shea},
        title = "{GizmoAnalysis: Read and analyze Gizmo simulations}",
 howpublished = {Astrophysics Source Code Library, record ascl:2002.015},
         year = 2020,
        month = feb,
          eid = {ascl:2002.015},
       adsurl = {https://ui.adsabs.harvard.edu/abs/2020ascl.soft02015W}
}

@ARTICLE{Dekel20,
       author = {{Dekel}, Avishai and {Ginzburg}, Omri and {Jiang}, Fangzhou and {Freundlich}, Jonathan and {Lapiner}, Sharon and {Ceverino}, Daniel and {Primack}, Joel},
        title = "{A mass threshold for galactic gas discs by spin flips}",
      journal = {\mnras},
         year = 2020,
        month = apr,
       volume = {493},
       number = {3},
        pages = {4126-4142},
          doi = {10.1093/mnras/staa470},
archivePrefix = {arXiv},
       eprint = {1912.08213},
 primaryClass = {astro-ph.GA},
       adsurl = {https://ui.adsabs.harvard.edu/abs/2020MNRAS.493.4126D}
}

@ARTICLE{Lazar20,
       author = {{Lazar}, Alexandres and {Bullock}, James S. and {Boylan-Kolchin}, Michael and {Chan}, T.~K. and {Hopkins}, Philip F. and {Graus}, Andrew S. and {Wetzel}, Andrew and {El-Badry}, Kareem and {Wheeler}, Coral and {Straight}, Maria C. and {Kere{\v{s}}}, Du{\v{s}}an and {Faucher-Gigu{\`e}re}, Claude-Andr{\'e} and {Fitts}, Alex and {Garrison-Kimmel}, Shea},
        title = "{A dark matter profile to model diverse feedback-induced core sizes of {\ensuremath{\Lambda}}CDM haloes}",
      journal = {\mnras},
         year = 2020,
        month = sep,
       volume = {497},
       number = {2},
        pages = {2393-2417},
          doi = {10.1093/mnras/staa2101},
archivePrefix = {arXiv},
       eprint = {2004.10817},
 primaryClass = {astro-ph.GA},
       adsurl = {https://ui.adsabs.harvard.edu/abs/2020MNRAS.497.2393L}
}

@ARTICLE{Freundlich20,
       author = {{Freundlich}, Jonathan and {Jiang}, Fangzhou and {Dekel}, Avishai and {Cornuault}, Nicolas and {Ginzburg}, Omry and {Koskas}, R{\'e}my and {Lapiner}, Sharon and {Dutton}, Aaron and {Macci{\`o}}, Andrea V.},
        title = "{The Dekel-Zhao profile: a mass-dependent dark-matter density profile with flexible inner slope and analytic potential, velocity dispersion, and lensing properties}",
      journal = {\mnras},
         year = 2020,
        month = dec,
       volume = {499},
       number = {2},
        pages = {2912-2933},
          doi = {10.1093/mnras/staa2790},
archivePrefix = {arXiv},
       eprint = {2004.08395},
 primaryClass = {astro-ph.GA},
       adsurl = {https://ui.adsabs.harvard.edu/abs/2020MNRAS.499.2912F}
}

@ARTICLE{Rohr22,
       author = {{Rohr}, Eric and {Feldmann}, Robert and {Bullock}, James S. and {{\c{C}}atmabacak}, Onur and {Boylan-Kolchin}, Michael and {Faucher-Gigu{\`e}re}, Claude-Andr{\'e} and {Kere{\v{s}}}, Du{\v{s}}an and {Liang}, Lichen and {Moreno}, Jorge and {Wetzel}, Andrew},
        title = "{The galaxy-halo size relation of low-mass galaxies in FIRE}",
      journal = {\mnras},
         year = 2022,
        month = mar,
       volume = {510},
       number = {3},
        pages = {3967-3985},
          doi = {10.1093/mnras/stab3625},
archivePrefix = {arXiv},
       eprint = {2112.05159},
 primaryClass = {astro-ph.GA},
       adsurl = {https://ui.adsabs.harvard.edu/abs/2022MNRAS.510.3967R}
}

@ARTICLE{RodriguezGomez22,
       author = {{Rodriguez-Gomez}, Vicente and {Genel}, Shy and {Fall}, S. Michael and {Pillepich}, Annalisa and {Huertas-Company}, Marc and {Nelson}, Dylan and {P{\'e}rez-Monta{\~n}o}, Luis Enrique and {Marinacci}, Federico and {Pakmor}, R{\"u}diger and {Springel}, Volker and {Vogelsberger}, Mark and {Hernquist}, Lars},
        title = "{Galactic angular momentum in the IllustrisTNG simulation -- I. Connection to morphology, halo spin, and black hole mass}",
      journal = {\mnras},
         year = 2022,
        month = jun,
       volume = {512},
       number = {4},
        pages = {5978-5994},
          doi = {10.1093/mnras/stac806},
archivePrefix = {arXiv},
       eprint = {2203.10098},
 primaryClass = {astro-ph.GA},
       adsurl = {https://ui.adsabs.harvard.edu/abs/2022MNRAS.512.5978R}
}

@ARTICLE{Sales22,
       author = {{Sales}, Laura V. and {Wetzel}, Andrew and {Fattahi}, Azadeh},
        title = "{Baryonic solutions and challenges for cosmological models of dwarf galaxies}",
      journal = {Nature Astronomy},
         year = 2022,
        month = jun,
       volume = {6},
        pages = {897-910},
          doi = {10.1038/s41550-022-01689-w},
archivePrefix = {arXiv},
       eprint = {2206.05295},
 primaryClass = {astro-ph.GA},
       adsurl = {https://ui.adsabs.harvard.edu/abs/2022NatAs...6..897S}
}

@ARTICLE{Yang23,
       author = {{Yang}, Hang and {Gao}, Liang and {Frenk}, Carlos S. and {Grand}, Robert J.~J. and {Guo}, Qi and {Liao}, Shihong and {Shao}, Shi},
        title = "{The galaxy size to halo spin relation of disc galaxies in cosmological hydrodynamical simulations}",
      journal = {\mnras},
         year = 2023,
        month = feb,
       volume = {518},
       number = {4},
        pages = {5253-5259},
          doi = {10.1093/mnras/stac3335},
archivePrefix = {arXiv},
       eprint = {2110.04434},
 primaryClass = {astro-ph.GA},
       adsurl = {https://ui.adsabs.harvard.edu/abs/2023MNRAS.518.5253Y}
}

@ARTICLE{Hopkins23,
       author = {{Hopkins}, Philip F. and {Wetzel}, Andrew and {Wheeler}, Coral and {Sanderson}, Robyn and {Grudi{\'c}}, Michael Y. and {Sameie}, Omid and {Boylan-Kolchin}, Michael and {Orr}, Matthew and {Ma}, Xiangcheng and {Faucher-Gigu{\`e}re}, Claude-Andr{\'e} and {Kere{\v{s}}}, Du{\v{s}}an and {Quataert}, Eliot and {Su}, Kung-Yi and {Moreno}, Jorge and {Feldmann}, Robert and {Bullock}, James S. and {Loebman}, Sarah R. and {Angl{\'e}s-Alc{\'a}zar}, Daniel and {Stern}, Jonathan and {Necib}, Lina and {Choban}, Caleb R. and {Hayward}, Christopher C.},
        title = "{FIRE-3: updated stellar evolution models, yields, and microphysics and fitting functions for applications in galaxy simulations}",
      journal = {\mnras},
         year = 2023,
        month = feb,
       volume = {519},
       number = {2},
        pages = {3154-3181},
          doi = {10.1093/mnras/stac3489},
archivePrefix = {arXiv},
       eprint = {2203.00040},
 primaryClass = {astro-ph.GA},
       adsurl = {https://ui.adsabs.harvard.edu/abs/2023MNRAS.519.3154H}
}

@ARTICLE{Lazar24,
       author = {{Lazar}, I. and {Kaviraj}, S. and {Watkins}, A.~E. and {Martin}, G. and {Bichang'a}, B. and {Jackson}, R.~A.},
        title = "{The morphological mix of dwarf galaxies in the nearby Universe}",
      journal = {\mnras},
         year = 2024,
        month = mar,
       volume = {529},
       number = {1},
        pages = {499-518},
          doi = {10.1093/mnras/stae510},
archivePrefix = {arXiv},
       eprint = {2402.12440},
 primaryClass = {astro-ph.GA},
       adsurl = {https://ui.adsabs.harvard.edu/abs/2024MNRAS.529..499L}
}

@ARTICLE{ManceraPina25,
       author = {{Mancera Pi{\~n}a}, Pavel E. and {Read}, Justin I. and {Kim}, Stacy and {Marasco}, Antonino and {Benavides}, Jos{\'e} A. and {Glowacki}, Marcin and {Pezzulli}, Gabriele and {Lagos}, Claudia del P.},
        title = "{The galaxy-halo connection of disc galaxies over six orders of magnitude in stellar mass}",
      journal = {\aap},
         year = 2025,
        month = jul,
       volume = {699},
          eid = {A311},
        pages = {A311},
          doi = {10.1051/0004-6361/202554381},
archivePrefix = {arXiv},
       eprint = {2505.22727},
 primaryClass = {astro-ph.GA},
       adsurl = {https://ui.adsabs.harvard.edu/abs/2025A&A...699A.311M}
}

@ARTICLE{Liang25a,
       author = {{Liang}, Jinning and {Jiang}, Fangzhou and {Mo}, Houjun and {Benson}, Andrew and {Dekel}, Avishai and {Tavron}, Noa and {Hopkins}, Philip F. and {Ho}, Luis C.},
        title = "{Connection between galaxy morphology and dark-matter halo structure I: a running threshold for thin discs and size predictors from the dark sector}",
      journal = {\mnras},
         year = 2025,
        month = aug,
       volume = {541},
       number = {3},
        pages = {2304-2323},
          doi = {10.1093/mnras/staf947},
archivePrefix = {arXiv},
       eprint = {2403.14749},
 primaryClass = {astro-ph.GA},
       adsurl = {https://ui.adsabs.harvard.edu/abs/2025MNRAS.541.2304L}
}

@ARTICLE{Benavides25,
       author = {{Benavides}, Jos{\'e} A. and {Sales}, Laura V. and {Wetzel}, Andrew and {Moreno}, Jorge and {Feldmann}, Robert and {Mercado}, Francisco J. and {Bullock}, James S. and {Hopkins}, Philip F. and {Faucher-Gigu{\'e}re}, Claude-Andr{\'e} and {Stern}, Jonathan and {Wheeler}, Coral and {Kere{\v{s}}}, Du{\v{s}}an},
        title = "{Disks no more: the morphology of low-mass simulated galaxies in FIREbox}",
      journal = {\mnras},
         year = 2025,
        month = oct,
       volume = {544},
       number = {4},
        pages = {4651-4664},
          doi = {10.1093/mnras/staf1847},
archivePrefix = {arXiv},
       eprint = {2508.00991},
 primaryClass = {astro-ph.GA},
       adsurl = {https://ui.adsabs.harvard.edu/abs/2025MNRAS.544.4651B}
}

@ARTICLE{Liang25b,
       author = {{Liang}, Jinning and {Jiang}, Fangzhou and {Mo}, Houjun and {Benson}, Andrew and {Hopkins}, Philip F. and {Dekel}, Avishai and {Ho}, Luis C.},
        title = "{Connection between galaxy morphology and dark-matter halo structure II: predicting disk structure from dark-matter halo properties}",
      journal = {arXiv e-prints},
         year = 2025,
        month = dec,
          eid = {arXiv:2512.13822},
        pages = {arXiv:2512.13822},
          doi = {10.48550/arXiv.2512.13822},
archivePrefix = {arXiv},
       eprint = {2512.13822},
 primaryClass = {astro-ph.GA},
       adsurl = {https://ui.adsabs.harvard.edu/abs/2025arXiv251213822L}
}



\appendix
\section{Quadratic response-surface analysis}
\label{app:rsm}
\subsection{Fitting overview}
In this Appendix we summarize the technical details of the quadratic response-surface model used in Section \ref{sec:rsm}. For each simulation in the $M_{\rm vir}=10^{11}\,{\rm M}_\odot$ grid we consider four halo predictors: concentration $c_2$, spin parameter $\lambda$, baryon fraction $f_{\rm b}$, and inner-slope proxy $s_1$. We work with transformed variables that reflect their natural scaling: $\log c_2$, $\log \lambda$, and $\log f_{\rm b}$ are used, while $s_1$ is kept on its original (linear) scale. For each transformed predictor $u_k$ we choose a reference value $u_{k,0}$ (the midpoint of the sampled range) and a half-range $\Delta u_k$, and define a dimensionless standardized coordinate
\begin{equation}
    z_k \equiv \frac{u_k - u_{k,0}}{\Delta u_k} ,
\end{equation}
so that $z_k \approx -1$ and $+1$ correspond to the low and high settings in the design, and $z_k=0$ corresponds to the fiducial model. The response variable is the natural logarithm of the 3D stellar half-mass radius, $y \equiv \log r_{1/2,\star}$, so multiplicative changes in size become additive shifts in $y$. In terms of the standardized predictors the quadratic response-surface model can be written as
\begin{equation}
    y = \beta_0 
      + \sum_{k} \beta_{1,k}\,z_k
      + \sum_{k} \beta_{2,k}\,z_k^2
      + \sum_{i<j} \beta_{ij}\,z_i z_j
      + \varepsilon ,
\end{equation}
where $\beta_0$ is the intercept, $\beta_{1,k}$ are the local linear slopes, $\beta_{2,k}$ describe curvature in each direction, $\beta_{ij}$ encode pairwise interactions, and $\varepsilon$ denotes the residuals. With four predictors this model contains 15 coefficients including the intercept. The chosen coding has the useful property that $\beta_0$ represents the mean $\log r_{1/2,\star}$ at the midpoints of the design, while the $\beta_{1,k}$ quantify the first-order sensitivity of galaxy size to each halo parameter around this reference point.

We estimate the coefficients by ordinary least squares (OLS), treating the simulation grid as a fixed experimental design. However, we do not rely on the usual OLS formula for uncertainties. Standard residual tests applied to an initial (non-robust) fit reveal significant departures from homoskedasticity: both the Breusch--Pagan and White tests reject the hypothesis of constant residual variance. In this regime we adopt the HC3 heteroskedasticity-consistent covariance estimator as implemented in \texttt{statsmodels}. HC3 downweights high-leverage points more strongly than simpler corrections (HC0/HC1) and yields more conservative small-sample inference. Throughout the paper we therefore quote OLS point estimates for the coefficients, but base all $t$- and $F$-tests, confidence intervals, and derived uncertainties on the HC3 covariance matrix. The design is well-conditioned: variance-inflation factors for all regression terms are close to unity (VIF $\lesssim 1.3$), indicating that collinearity among the standardized predictors is not a concern.

For the full $N=132$ sample at $M_{\rm vir}=10^{11}\,{\rm M}_\odot$, the quadratic response surface provides a good global description of the size response. The coefficient of determination is $R^2 \simeq 0.80$ (adjusted $R^2 \simeq 0.78$), with an overall $F$-statistic $F_{14,117} \simeq 36.4$ and $p \simeq 5\times10^{-36}$. The root-mean-square residual in $\log r_{1/2,\star}$ is $\simeq 0.37$, corresponding to a $1\sigma$ scatter of $\simeq 0.16$\,dex, or a multiplicative factor of $\sim 1.5$ in physical radius. Only a handful of runs have large studentized residuals or are flagged as influential by Cook's-distance and leverage thresholds. Re-fitting the model after removing these high-influence cases increases $R^2$ to $\simeq 0.86$ and leaves the main-effect curves unchanged within uncertainties, indicating that our conclusions are not sensitive to a small subset of outlying realisations.

\subsection{Importance proxies from coefficient-block tests}

Table~\ref{tab:importance} reports two importance proxies derived from the response-surface model. Both are based on joint Wald tests on blocks of coefficients using the HC3 covariance matrix. For a block test with statistic $F$, numerator degrees of freedom $d$, and residual degrees of freedom $\nu$, we convert the result into an effect-size proxy on a 0--1 scale via
\begin{equation}
    R^2_{\rm partial} \equiv \frac{d\,F}{d\,F + \nu}.
\end{equation}
This quantity is closely related to the ``partial $\eta^2$'' used in analysis-of-variance settings; in our context it provides a convenient, approximate measure of the additional (conditional) variance explained by the tested block given the rest of the model. Because it is conditional on the other terms in the regression, it is not additive across factors and should be interpreted comparatively.

For the main-effect proxy we test the joint contribution of the linear and quadratic terms of each factor, $(\beta_{1,k},\beta_{2,k})$ ($d=2$). This yields large and highly significant partial $R^2$ values for $c_2$, $\lambda$, and $s_1$ (all $\simeq 0.46$--0.49 with $p_F\ll 10^{-6}$), and a much smaller, non-significant value for $f_{\rm b}$ (partial $R^2\simeq 0.03$ with $p_F\simeq 0.14$). For the total-effect proxy we expand the tested block to include the linear and quadratic terms \emph{plus} all interactions involving the factor (five coefficients in total; $d=5$). In this case, $c_2$, $\lambda$, and $s_1$ remain in a high-importance tier (partial $R^2\simeq 0.52$--0.58), while the importance of $f_{\rm b}$ increases substantially (partial $R^2\simeq 0.21$) owing to its coupling with spin. This is consistent with Fig.\ref{fig:rsm_interaction}, where the \((f_{\rm b},\lambda)\) panel shows the clearest non-separable behavior.

\section{Random-forest training and feature-importance definitions}
\label{app:rf}

\subsection{Model setup and global hyper-parameter tuning}
We model the mapping from halo parameters to galaxy size using a random-forest regressor (RF; \citealt{Breiman01}). Each RF consists of an ensemble of decision trees trained on bootstrap resamples of the training set, and predictions are obtained by averaging over trees. Because tree-based models are invariant to monotonic feature rescalings and do not require feature standardization, we train directly on the physical input parameters $(c_2, s_1, \lambda, f_{\rm b})$.

We select the RF hyper-parameters with a single global \texttt{GridSearchCV} pass on the full $N=132$ sample, using 5-fold CV and the negative mean-squared error as the scoring function. The explored grid is
\[
\begin{aligned}
&n_{\rm est} \in \{300,600,800,1000,1200,1500\},\\
&{\rm max\_depth} \in \{\texttt{None},5,10,15\},\\
&{\rm min\_samples\_leaf} \in \{1,2,3,5\},\\
&{\rm max\_features} \in \{\texttt{sqrt},1.0\}.
\end{aligned}
\]

The best-performing configuration is
\[
\begin{aligned}
&n_{\rm est} = 1000,\\ 
&{\rm max\_depth}=10,\\
&{\rm min\_samples\_leaf} = 1,\\
&{\rm max\_features}=\texttt{sqrt}.
\end{aligned}
\]

with bootstrap sampling enabled and \texttt{oob\_score=True}. These hyper-parameters are then held fixed for all subsequent resampling experiments to avoid confounding the stability of feature importances with repeated re-tuning.

\subsection{Repeated cross-validation protocol and model diagnostics}
To assess the robustness of importance rankings to the limited sample size and to the particular train/validation partition, we perform repeated $K$-fold cross-validation with $K=5$ and $N_{\rm rep}=20$, yielding 100 train/validation splits. Each split trains on $\simeq 4/5$ of the sample ($\sim$105--106 runs) and evaluates on the held-out fold ($\sim$26--27 runs). We report standard performance diagnostics to ensure that the RF captures genuine signal: over the 100 splits, the out-of-fold coefficient of determination is $\langle R^2_{\rm valid}\rangle\simeq 0.64$ with scatter $\sigma(R^2_{\rm valid})\simeq 0.19$. The out-of-bag score, computed internally on OOB samples within each training fold, is comparably high and substantially more stable, $\langle R^2_{\rm OOB}\rangle\simeq 0.64$ with $\sigma(R^2_{\rm OOB})\simeq 0.05$. This indicates that the trained forests are not dominated by pathological overfitting in individual splits, and that importance estimates can be meaningfully aggregated across repeats.

\subsection{Feature-importance estimators}
We quantify feature importance using four complementary estimators that probe different notions of “importance”:

\begin{enumerate}[leftmargin=*,labelindent=0pt,labelsep=0.5em]
\item \textbf{Mean decrease in impurity (MDI).} We use the standard impurity-based importance returned by the RF implementation, which measures the average reduction in within-node variance attributable to each feature, accumulated over all splits and trees.

\item \textbf{Out-of-fold (OOF) permutation importance.} For each CV split, we evaluate the baseline validation-set mean-squared error ${\rm MSE}_{\rm valid}$, then randomly permute a single feature column $x_j$ in the validation set (holding all other columns fixed) and re-evaluate ${\rm MSE}^{\pi_j}_{\rm valid}$. The permutation importance is the performance degradation
\begin{equation}
I^{\rm perm}_j \equiv {\rm MSE}^{\pi_j}_{\rm valid} - {\rm MSE}_{\rm valid},
\end{equation}
averaged over repeated shuffles. This estimator directly measures how much predictive accuracy is lost when the information in $x_j$ is destroyed out of sample.

\item \textbf{Out-of-bag (OOB) permutation importance.} For each tree in the forest, we identify its OOB subset (samples not included in the bootstrap draw for that tree). We compute a baseline OOB error for that tree, then permute feature $x_j$ within the OOB subset and recompute the error. The per-tree degradation is averaged over trees (and repeated permutations) to obtain an OOB-based permutation importance. Compared to OOF permutation, this avoids using the held-out validation fold and instead exploits the bootstrap structure of the RF.

\item \textbf{SHAP-based global importance.} We compute Shapley-style additive explanations (SHAP) for tree ensembles using a TreeExplainer approach (e.g. \citealt{Lundberg17}). For each split we summarise global importance by the mean absolute SHAP value of each feature over the validation fold, $\langle |{\rm SHAP}_j|\rangle$, which measures the typical magnitude of the feature’s contribution to the model output.
\end{enumerate}

Because these estimators live on different scales (and permutation scores can occasionally be slightly negative in finite samples), we convert each split-level importance vector into a normalised score that is comparable within a method:
\begin{equation}
\tilde I_j \equiv \frac{\max(I_j,0)}{\sum_k \max(I_k,0)} ,
\end{equation}
so that $\sum_k \tilde I_k=1$ for each estimator and split.

\subsection{Aggregation, composite score, and resulting hierarchy}
Table~\ref{tab:importance} reports, for each RF estimator, the mean of $\tilde I_j$ over the 100 repeated-CV splits. We also define a \textbf{composite} RF importance as the arithmetic mean of the four normalised estimators (MDI, OOF permutation, OOB permutation, and SHAP) for each split, followed by averaging over splits. The composite scores show a stable ordering
\[
c_2 > (\lambda \simeq s_1) > f_{\rm b},
\]
with $c_2$ consistently dominant and $f_{\rm b}$ consistently least important. The robustness is also evident in rank stability: $c_2$ is ranked first in the overwhelming majority of resamples (depending on estimator, $\sim$80--100\% of splits), while $f_{\rm b}$ is almost always ranked last. The relative ordering between $\lambda$ and $s_1$ varies mildly with the estimator (permutation-based metrics tend to favour $\lambda$, whereas MDI/SHAP slightly favour $s_1$), motivating our conservative conclusion that they occupy a shared intermediate tier.

\section{Transplanting the Mo98 formalism to the Dekel--Zhao halo profile}
\label{app:mmw_dz}

\subsection{A root-finding solver for the Mo98 disc scale length}
\label{app:mmw_root}

The classical disc-formation model of \citet[][hereafter Mo98]{Mo98} assumes that a fraction $m_{\rm d}$ of the halo virial mass $M_{\rm vir}$ settles into a rotationally supported exponential disc and retains a fraction $j_{\rm d}$ of the halo angular momentum.\footnote{In the main text we often use the retention factor $f_j\equiv j_{\rm d}/m_{\rm d}$.}
The disc surface density is
\begin{equation}
\Sigma(R)=\frac{M_{\rm d}}{2\pi R_{\rm d}^2}\exp\!\left(-\frac{R}{R_{\rm d}}\right),
\qquad
M_{\rm d}=m_{\rm d}\,M_{\rm vir},
\label{eq:app_Sigma}
\end{equation}
and the disc angular momentum for a given circular-velocity curve $V_c(R)$ is
\begin{equation}
\begin{aligned}
&J_{\rm d}(R_{\rm d})=2\pi\int_0^\infty \Sigma(R)\,V_c(R;R_{\rm d})\,R^2\,{\rm d}R\\
       &= M_{\rm d} R_{\rm d} \int_0^\infty e^{-u}u^2\,V_c(uR_{\rm d};R_{\rm d})\,{\rm d}u,
\end{aligned}
\label{eq:app_Jd}
\end{equation}
where $u\equiv R/R_{\rm d}$. The target angular momentum is $J_{\rm d}=j_{\rm d} J_{\rm h}$, with the halo angular momentum $J_{\rm h}$ obtained from the Peebles spin parameter (Mo98 eq.~9),
\begin{equation}
\lambda \equiv \frac{J_{\rm h}\,|E_{\rm h}|^{1/2}}{G\,M_{\rm vir}^{5/2}}
\quad \Longrightarrow \quad
J_{\rm h}=\lambda\,\frac{G\,M_{\rm vir}^{5/2}}{|E_{\rm h}|^{1/2}}.
\label{eq:app_spin}
\end{equation}
Writing the halo total energy in the standard form
\begin{equation}
E_{\rm h} = -\frac{1}{2}\,f_{\rm h}\,M_{\rm vir}V_{\rm vir}^2
          = -\frac{G M_{\rm vir}^2}{2R_{\rm vir}}\,f_{\rm h},
\qquad
V_{\rm vir}^2\equiv \frac{G M_{\rm vir}}{R_{\rm vir}},
\label{eq:app_energy_def}
\end{equation}
one can rewrite eq.~\eqref{eq:app_spin} as
\begin{equation}
J_{\rm h}=\lambda\,M_{\rm vir}R_{\rm vir}V_{\rm vir}\sqrt{\frac{2}{f_{\rm h}}}.
\label{eq:app_Jhalo_simplified}
\end{equation}
For an NFW halo, Mo98 give $E_{\rm h}=-(G M_{\rm vir}^2/2R_{\rm vir})\,f_c$ (Mo98 eq.~22), with
\begin{equation}
f_c(c)=\frac{c}{2}\,
\frac{1-(1+c)^{-2}-2\ln(1+c)/(1+c)}
{\left[c/(1+c)-\ln(1+c)\right]^2}
\label{eq:app_fc_nfw}
\end{equation}
(Mo98 eq.~23), where $c$ is the concentration.

In Mo98 the disc scale length is obtained by rewriting the condition $J_{\rm d}=j_{\rm d}J_{\rm h}$ into a fixed-point form (their eqs.~28--29) and iterating. However, the iterative scheme is not essential: for any specified initial halo profile (NFW in Mo98; DZ in our work), the adiabatic-contraction calculation always starts from the same unperturbed halo, so the mapping $R_{\rm d}\mapsto J_{\rm d}(R_{\rm d})$ is deterministic. One may therefore solve for $R_{\rm d}$ directly as a one-dimensional root-finding problem,
\begin{equation}
\mathcal{F}(R_{\rm d})\equiv J_{\rm d}(R_{\rm d})-j_{\rm d} J_{\rm h}=0,
\label{eq:app_Rd_root}
\end{equation}
where $J_{\rm d}(R_{\rm d})$ is evaluated by direct numerical integration of eq.~\eqref{eq:app_Jd}.

Evaluating $V_c(R;R_{\rm d})$ follows Mo98: the halo responds adiabatically to the slow assembly of the disc, conserving $rM(r)$ for each shell,
\begin{equation}
r_i\,M_i(r_i)=r\,M_f(r),
\label{eq:app_ac}
\end{equation}
where $M_i(r)$ is the initial halo enclosed mass profile, and
\begin{equation}
M_f(r)=M_{\rm d}(r)+M_{\rm h}(r),
\qquad
M_{\rm h}(r)=(1-m_{\rm d})\,M_i(r_i).
\label{eq:app_Mf}
\end{equation}
The cumulative mass of an exponential disc is (Mo98 eq.~26)
\begin{equation}
M_{\rm d}(r)=M_{\rm d}\left[1-\left(1+\frac{r}{R_{\rm d}}\right)e^{-r/R_{\rm d}}\right].
\label{eq:app_Md}
\end{equation}
The total rotation curve is then
\begin{equation}
V_c^2(r)=V_{\rm d}^2(r)+V_{\rm h}^2(r),
\qquad
V_{\rm h}^2(r)=\frac{G M_{\rm h}(r)}{r},
\label{eq:app_Vc}
\end{equation}
with the thin exponential-disc contribution
\begin{equation}
V_{\rm d}^2(r)=\frac{G M_{\rm d}}{2R_{\rm d}}\,y^2\Big[I_0(y)K_0(y)-I_1(y)K_1(y)\Big],
\qquad
y\equiv \frac{r}{2R_{\rm d}},
\label{eq:app_Vd}
\end{equation}
where $I_n$ and $K_n$ are modified Bessel functions. In practice, eq.~\eqref{eq:app_Rd_root} can be solved robustly with a bracketing method (e.g. Brent's algorithm), and is typically fast on modern hardware while being conceptually more transparent than an explicit fixed-point iteration.

\subsection{Dekel-Zhao haloes: \texorpdfstring{$M(r)$}{M(r)} and \texorpdfstring{$f_{\rm DZ}$}{f\_DZ}}
\label{app:dz}

To transplant the Mo98 machinery to a Dekel--Zhao (DZ) halo \citep{Freundlich20}, only two halo-specific ingredients are required:
(i) the initial enclosed mass profile $M_i(r)$ entering the adiabatic-contraction step (eq.~\ref{eq:app_ac}); and
(ii) the energy factor $f_{\rm h}$ entering the spin--angular-momentum conversion (eq.~\ref{eq:app_Jhalo_simplified}).

The DZ density profile is (Freundlich et al.\ 2020, eq.~11)
\begin{equation}
\rho(r)=\frac{\rho_c}{x^a\left(1+x^{1/2}\right)^{2(3.5-a)}},
\qquad
x\equiv \frac{r}{r_c},
\qquad
c\equiv \frac{R_{\rm vir}}{r_c},
\label{eq:app_dz_rho}
\end{equation}
with inner-slope parameter $a$ and concentration $c$.
It is convenient to define the compact variable
\begin{equation}
\chi(x)\equiv \frac{x^{1/2}}{1+x^{1/2}},
\qquad
\chi_c\equiv \chi(c)=\frac{c^{1/2}}{1+c^{1/2}}.
\label{eq:app_chi}
\end{equation}
Then the enclosed mass takes the simple closed form
\begin{equation}
M_i(r)=M_{\rm vir}\left(\frac{\chi}{\chi_c}\right)^{6-2a}.
\label{eq:app_M_dz}
\end{equation}

For the total energy, we define the DZ analogue of the NFW energy factor by
\begin{equation}
E_{\rm h}\equiv -\frac{G M_{\rm vir}^2}{2R_{\rm vir}}\,f_{\rm DZ}(a,c)
              =-\frac{1}{2}\,f_{\rm DZ}(a,c)\,M_{\rm vir}V_{\rm vir}^2.
\label{eq:app_E_dz}
\end{equation}
Assuming circular orbits for the virialised halo and using the virial theorem, one obtains
\begin{equation}
f_{\rm DZ}(a,c)=\frac{2(3-a)c}{\chi_c^{12-4a}}\,B\!\left(\chi_c;10-4a,3\right),
\label{eq:app_fdz}
\end{equation}
where $B(z;p,q)$ is the (unnormalised) incomplete beta function,
\begin{equation}
B(z;p,q)\equiv \int_0^z t^{p-1}(1-t)^{q-1}\,{\rm d}t,
\label{eq:app_incbeta}
\end{equation}
and the integral converges for the parameter range relevant here (in particular $a<2.5$ so that $p=10-4a>0$).

With eqs.~\eqref{eq:app_M_dz} and \eqref{eq:app_fdz}, the Mo98 procedure carries over verbatim:
in the contraction step, replace the NFW $M_i(r)$ by the DZ mass profile;
in the spin conversion, replace $f_c(c)$ by $f_{\rm DZ}(a,c)$.
The disc scale length $R_{\rm d}$ is then obtained by solving the same root condition, eq.~\eqref{eq:app_Rd_root}.


\bsp	
\label{lastpage}
\end{document}